\documentclass[12pt]{iopart}

\usepackage{graphicx}
\usepackage{multirow}
\usepackage{subcaption}
\expandafter\let\csname equation*\endcsname\relax
\expandafter\let\csname endequation*\endcsname\relax
\usepackage{amsmath}
\usepackage{xcolor}
\usepackage{framed} 
\colorlet{shadecolor}{red!10}

\begin{document}

\title[A Criticism of Interpretable ML for Intrusion Detection]{A Critical Assessment of Interpretable and Explainable Machine Learning for Intrusion Detection} 

\author{Omer Subasi, Johnathan Cree, Joseph Manzano, Elena Peterson}

\address{Pacific Northwest National Laboratory, 902 Battelle Blvd, Richland, 99354, WA, USA}
\ead{omer.subasi@pnnl.gov, johnathan.cree@pnnl.gov, joseph.manzano@pnnl.gov, elena@pnnl.gov}

\begin{abstract}
Interpretable and explainable Machine Learning (ML) refers to methods and models
that explain how and why a model makes a prediction and describe a model's behaviour
in human understandable terms. There has been a large number of studies in
interpretable and explainable ML for cybersecurity, in particular, for intrusion detection.
Many of these studies have significant amount of overlapping and repeated evaluations and analysis.
At the same time, these studies overlook crucial model, data, learning process, and utility
related issues and many times completely
disregard them. These issues include the use of overly complex and opaque ML models, 
unaccounted data imbalances and correlated features,
inconsistent influential features across different explanation methods, 
the inconsistencies stemming from the constituents of a learning process, and
the implausible utility of explanations. 

Therefore, in this work, we empirically demonstrate the aforementioned issues, analyze them and propose 
practical solutions in the context of \textit{feature-based model explanations}. 
Specifically, we advise avoiding complex opaque models such as Deep Neural Networks (DNNs) 
and instead using interpretable ML models such as Decision Trees (DTs)
as the available intrusion datasets are not difficult for such interpretable models to classify successfully.
Then, we bring attention to the binary classification metrics 
such as Matthews Correlation Coefficient (MCC) which are well-suited for imbalanced datasets.
Moreover, we find that feature-based model explanations are most often inconsistent across different settings.
In this respect, to further gauge the extent of inconsistencies,
we introduce the notion of \textit{cross explanations} which corroborates 
that the features that are determined to be impactful 
by one explanation method most often differ from those by another method.
Furthermore, we show that strongly correlated data features and the constituents of a learning process, such as 
hyper-parameters and the optimization routine, become yet another source of inconsistent explanations.
Finally, we discuss the utility of feature-based explanations. 

In our empirical study, we use two real-world intrusion datasets, two interpretable classification models, namely, linear
Ridge and DT classifiers, and a DNN-based classifier together with 
SHapley Additive exPlanations (SHAPs) and Permutation feature Importances (PIs) 
external explanation methods.
\end{abstract}


\maketitle

\section{Introduction}
Artificial Intelligence (AI) and Machine Learning (ML) 
have achieved significant advances in the last decade.
The breathtaking progress in Deep Learning (DL) 
has unleashed powerful Deep Neural Networks (DNNs) solving complex tasks in 
computer vision, natural language processing, speech recognition, 
and many others \cite{Goodfellow2016}.
Naturally, the adoption of these methods and models in cybersecurity research has greatly
expanded \cite{aisurvey1}. 
Researchers have applied and evaluated various models and algorithms
to predict, detect, and mitigate cyber-attacks \cite{aisurvey1,macas2022survey}.

With ever-widening adoption of DNNs, understanding and interpreting their predictions
has become a significant problem \cite{molnar2020interpretable}. 
This is because DNNs are notoriously hard to understand since they have an  
opaque, often complex, internal structure. For this reason, they are also called ``black-box" methods. 
As DNNs grow in size, explaining how they make their predictions
becomes harder and nontrivial. Consequently, this very problem of explainability and interpretability of 
ML/DL models and methods carries to cybersecurity research as their
adoption becomes widespread in cybersecurity community \cite{surveyaicyber}.

To interpret and explain the outputs of ML models, 
there has been a large number of studies \cite{surveyaicyber} 
and, in particular, for intrusion detection \cite{moustafa2023explainable}. 
The vast majority of the existing studies on interpretable and explainable intrusion detection
evaluates ML/DL-based detection methods with a widely-used cyber-attack (intrusion) dataset and
some \textit{feature-based explanation} method. 
These studies inadvertently consist of significantly overlapping repeated work \cite{moustafa2023explainable}. 
As a result, they commonly fail to consider and address certain ML-related issues. 
After a careful review of the state of the art literature, 
we see that existing studies often do not investigate  
the selection of appropriate ML classification models by 
taking the complexity of existing datasets into account. Similarly,
they do not examine class imbalances or strongly correlated features.
We also see that although \textit{feature-based model explanations} are largely inconsistent, 
the instigating aleatoric (stochastic) uncertainties are not scrutinized. 
Moreover, the literature has not explored 
the impact of the constituents (elements) of a learning process, 
such as hyper-parameters and optimization methods.
Lastly, we see that existing studies do not assess
the utility of feature-based model explanations. 

Therefore, in this work, we focus on these critical, however unattended, 
problems in explainability and interpretability of ML-based
intrusion detection methods. First, we empirically show that low-cost
interpretable methods such as Decision Trees (DTs) are successful
in the classification of intrusions in the existing cyber-attack datasets.
Second, our examination of these datasets reveals a significantly
high level of class imbalance which is also noted by \cite{dodge}. In consequence, 
we advocate for the adaption of balanced accuracy (BA) and 
\textit{Matthews Correlation Coefficient (MCC)} \cite{MATTHEWS1975442}, which 
is considered as one of the most reliable and accurate classification metrics 
\cite{Chicco2021-tv, Chicco2023-la}.

Furthermore, we introduce the notion of \textit{cross-explanations}: 
Given a fixed dataset, we first compute the most influential features 
that impact the predictions of an ML-based intrusion/attack classifier as
determined by an external explanation method or the ML classifier itself.
We then evaluate the classification performance of another (type of) classifier by only using these features.
This is to see if these features are \textit{transferable}. That is, 
whether the features that are found to be impactful for one ML-based classifier
are similarly impactful on the performance of another classifier. 
As a result, cross-explanations
provide a new way of probing if the explanations, i.e., the most impactful features, are consistent
across different classifiers and settings.

Finally, we analyze how feature correlations and the constituents of a 
learning process alter the explanations. 
Our results indicate that strongly correlated features and these constituents, e.g., hyperparameters and
the optimization routine, become sources of aleatoric uncertainty and thus cause 
unstable and contradictory explanations.

\textit{Overall, the experimental evidence 
clearly suggests that a vast majority of the feature-based explanations, 
regardless of being obtained intrinsically or externally,
has little value in terms of practical applicability due to being inconsistent, unstable, and unreliable.
Coupled with the inability to offer any actionable course, we conclude that
these explanations do not provide any tangible utility in cybersecurity applications and practices.
Thus, research studies should prioritize the approaches that are not based on feature importance such as 
prototype-based methods and counterfactual explanations.
}

To summarize our main contributions,
\begin{itemize}
    \item We conduct a thorough analysis of \textit{feature-based} explainable and interpretable ML 
    for intrusion detection in which we investigate the unattended problems in cybersecurity research. 
    \item We empirically show that DTs successfully classify the intrusions in the 
    existing intrusion datasets. This signifies that complex DNNs are not needed. 
    \item We bring attention to class imbalance that exists in many cybersecurity datasets and advise the adaption of
     BA and MCC \cite{MATTHEWS1975442}.
    \item We mathematically analyze how a misalignment between a model and its feature-based explanation can lead to incorrect conclusions.
    \item We experimentally substantiate that there are pervasive inconsistencies among the most important features affecting the predictions of an ML model due to aleatoric uncertainty. 
    \item We introduce the method of \textit{cross-explanations} as a novel way to 
    assess the reliability and consistency of explanations.
    \item  We study the effect of feature correlations, hyper-parameters values 
    and the type of an optimization routine on the consistency of the most impactful features.
    \item We use SHAP \cite{shapley}, Permutation feature Importance (PI) \cite{breiman2001random}, 
    and classifier-intrinsic feature coefficients and importances with the newly available 5G dataset \cite{5gdataset} and 
     the UDBLag dataset \cite{internetdataset}. 
     Throughout our study, we highlight key results and offer specific guidelines.
\end{itemize}

In the next section, we provide the background and related work. In Section \ref{experimentalsetup},
we detail the experimental setup. In Section \ref{eval}, we present our results and in-depth analysis. Finally,
Section \ref{conclusion} concludes the study.

\section{Background and Related Work}
\label{background}
In this section, we first provide the background for explainable and interpretable ML. 
Second, we summarize the state of the art external explanation methods.
Then, we explain the task of (binary) classification and several metrics to evaluate the classification 
performance of an ML model. 
Finally, we discuss the related work.

\subsection{Interpretable and Explainable ML}
Interpretable and explainable ML \cite{molnar2020interpretable} is a research field 
whose main purpose is to reason,
interpret, and explain why and how ML models make the predictions they make.
It often attempts to quantify the impact of data features on a model's prediction and to correctly 
understand and interpret the model's outputs. 
Interpretable and explainable ML are crucial and often required in certain domains and fields, 
such as healthcare, finance and cybersecurity.
While the notions of interpretability and
explainability are often used interchangeably, there is some difference between them.
Interpretability refers to the extent to which the inner workings of an ML
method can be understood \cite{molnar2020interpretable}. 
It leads to understanding how a model's parameters and the input data features determine the 
model's outputs. To put differently, interpretability is the extent to 
which a cause and effect can be observed in an ML system. 
Explainability, on the other hand, refers to the extent to which the inner workings of
an ML model can be explained in human terms.  

Interpretable ML models can be understood on their own. These models 
do not require external methods. Examples are linear models and
Decision Trees (DTs) \cite{breiman2017classification}. 
DTs are highly interpretable since a DT's prediction can be understood by
simply following the path from the root to the leaf node and logically conjoining
the node conditions. The  
Feature Importances (FIs) of Decision Trees (DTs)
are intrinsically computed 
by the reduction a feature makes in the criterion, such as Gini Index and Entropy,
that is used to select the splits in a DT.
FIs can be viewed as self-feedback of DTs and 
they provide another means to understand and interpret DTs. 
That is, DTs provide two forms of explanations: i)
the tree itself - from which boolean logic rules are extracted - and ii) the FIs.
As for linear models, the feature coefficients (FCs)  
can be used to understand which features are more influential than others on a model's output. 
FCs are categorically different than DTs' FIs.
They are conditional in the sense that they quantify the weight of
a feature assuming that all other features are fixed. To be on the same footing
with FIs scores, FCs need to be corrected such that all features have the same unit.
Standardization is an approximate way of achieving a common unit.
In our analysis, we choose to study Ridge classifiers which are linear models with 
regularization that is quadratic in FCs. 
Regularization can prevent over-fitting and alleviate the problems arising due to strongly correlated features.

\begin{figure}[t]
    \centering
    \includegraphics[width=1\linewidth]{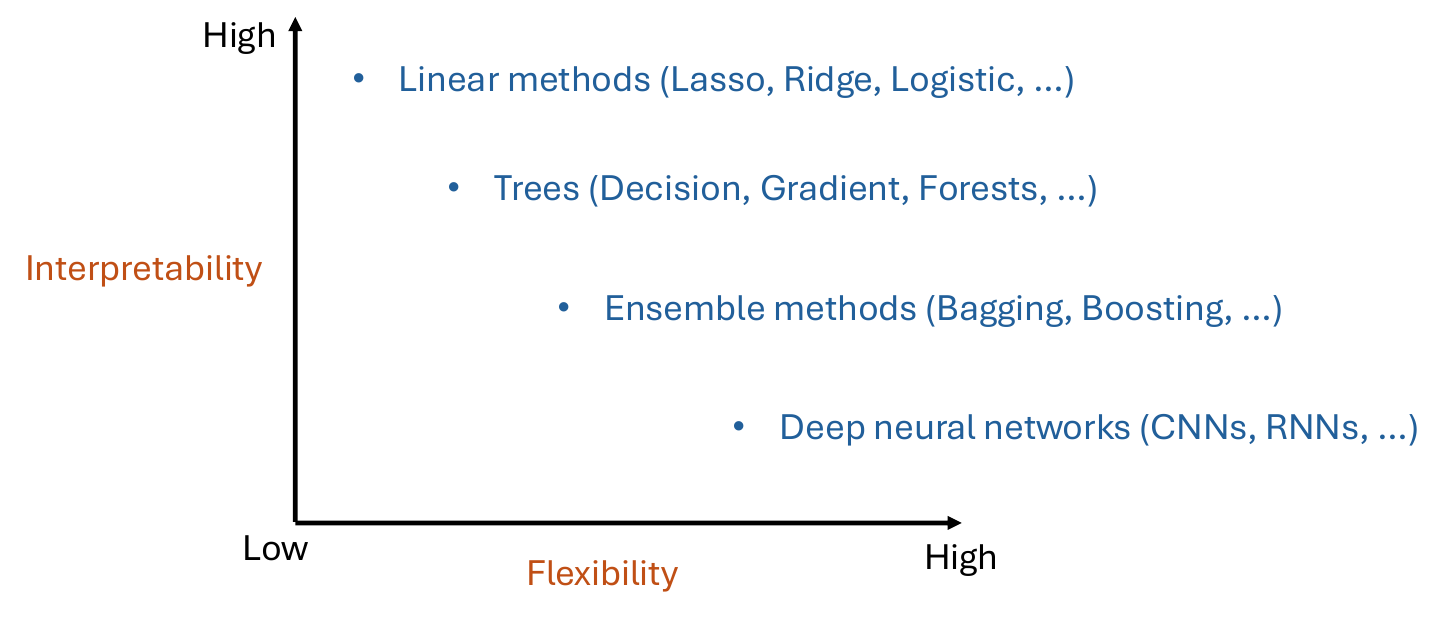}
    \caption{Interpretability vs. flexibility of ML models. 
    Note that the relationship shown here is a qualitative  overview of ML models.}
    \label{fig:ml_interpretablity_performance}
\end{figure}

Compared to interpretable ML models, explainable models are too complex for a human to 
understand on their own. These models also called ``black-box" and ``opaque" models. 
They require external explanation methods to be understood by humans.
DNNs are the prototypical examples of black-box models.

Considering the overall relationship between model interpretability and flexibility,
Figure \ref{fig:ml_interpretablity_performance} shows how the interpretability of ML
models varies with respect to the flexibility of those models in general. We see that linear 
and tree-based models are more interpretable than ensemble models and DNNs. 
This comes at the cost of relatively lesser flexibility and potentially lower performance. 
However, our results in Section
\ref{eval} show that for the intrusion detection datasets we evaluated, 
DTs perform as successfully as DNNs. 

\subsection{External Explanation Methods}
Existing external explanation methods and techniques can be classified 
as local and global methods. Local methods explain the specific outputs 
of specific inputs. That is, they focus on individual input-output pairs.
Local Interpretable Model-agnostic Explanations (LIME) \cite{lime} and 
Individual Conditional Expectations \cite{goldstein2015peeking}
are well-known local methods. 
In particular, LIME explains an ML model 
by assuming that the model is approximately and locally linear around 
the specific input instance that is under consideration. The explanations are then computed by the
weights of the linear approximation.
Considering global explanation methods, 
SHapley Additive exPlanations (SHAP) \cite{shapley} is one of the most popular
approaches that quantifies the contribution of input features to the model's output by approximating the
performance difference between including and not including the feature in the training.
Explanation methods can also be classified as model-agnostic and model-specific
methods. Model-agnostic methods consider any ML model as a black-box model, that is, the internal
workings of the model are not known or understood. Examples are LIME and SHAP as both assume no knowledge of 
a model's internal working and they work based on input-output pairs.
Permutation feature Importance (PI) \cite{breiman2001random} is another (global) 
model-agnostic method that quantifies the impact of a feature
on a model's output by computing the change 
in the model's performance while shuffling the feature's values. On the other hand,
model-specific methods are tailored for specific ML models. As an example,  
Montavon et. al. \cite{montavon2019layer} propose a model-specific method for DNNs that is built on the loss function's gradients.

Some external explanation methods are prototype-based  \cite{Prototype, Bayesianproto, prototypes2018, prototypenauta2021neural}. These methods aim to construct a set of prototypes which is representative of a dataset, e.g., a set of images of the ten digits (0-9). 
Depending on the context, a prototype can be an input data instance from a dataset, an observation that is very close to a data instance or a combination of several data instances. In classification tasks,
the class of a data instance is typically determined to be the class of the 
``closest" or ``the most similar" prototype where closeness or similarity is defined by a distance metric. 

Other than the aforementioned methods,
counterfactual explanations are the sets of features that need to change 
to achieve a desired model prediction \cite{karimi2020model, wachter2017counterfactual}, while anchors are sufficient local conditions consisting of if-then rules \cite{Anchors}.
Closely related to all these feature-based explanations, 
feature selection and dimensionality reduction algorithms \cite{subasi2024analysis} also provide the set of 
the most ``important" features. The definition of importance varies across different contexts and methods.

External explanation methods unsurprisingly have limitations.
For instance, SHAPs are exponential costly in the number of features and therefore heuristic algorithms are required. 
LIME, on the other hand, can be unstable and there is no guarantee that the
linearity assumption will hold in all circumstances. PIs fail when there are strongly correlated features or the trained model itself performs poorly.

\subsection{Binary Classification and Related Metrics}
Machine classification refers to the problem of categorizing a data instance into 
the predefined categories. As a special case, binary classification
assumes the number of categories or classes is two, where the classes can 1 and 0, or
true and false, or cats and dogs, or attack and no attack.
Binary classification is typically performed by supervised learning methods such as
linear classifiers, e.g., Logistic Regression and Ridge, nearest neighbors methods, support vector
machines (SVMs), Decision Trees (DTs), random forests, and deep neural networks (DNNs).
There are many performance metrics for binary classification and they are all based on the standard 2x2 binary classification confusion matrix as shown in Table \ref{tab:confusionmat}.
Table \ref{tab:metric_defs} defines the metrics we use in our study.
\begin{table}[ht]
\centering
\caption{Confusion Matrix}
\begin{tabular}{|c|c|c|c|} 
\hline
& & \multicolumn{2}{c|}{Predicted} \\ \hline
& & Positive (P) & Negative (N) \\ 
\hline
\multirow{2}{3em}{Actual} & Positive (P) & True Positive (TP) & False Negative (FN) \\  
& Negative (N) & False Positive (FP) & True Negative (TN)  \\ \hline
\end{tabular}
\label{tab:confusionmat}
\end{table}
\begin{table}[ht]
\centering
\caption{Definition of Classification Metrics}
\begin{tabular}{|l|l|}
\hline
Metric & Definition \\ \hline
Accuracy & $\frac{TP + TN}{TP + TN + FP + FN}$ \\ \hline
 Balanced Accuracy (BA) & $ \frac{1}{2} (\frac{TP}{TP + FN} + \frac{FP}{FP + TN})$ \\ \hline
F1 Score & $\frac{2TP}{2TP + FP + FN}$ \\ \hline
Precision & $\frac{TP}{TP + FP}$ \\ \hline
Recall  & $\frac{TP}{TP + FN}$ \\ \hline
Matthews Correlation Coefficient (MCC) & $\frac{TP \times TN - FP \times FN}{\sqrt{(TP + FP) \times (TP + FN)  \times (FP + TN) \times (TN + FN))}}$ \\ \hline
\end{tabular}
\label{tab:metric_defs}
\end{table}
These metrics include standard metrics such as accuracy, precision, recall and F1 score. 
In contrast to much of the existing literature in cybersecurity and ML in general,
we include the lesser known  
metrics of Balanced Accuracy (BA) and Matthews Correlation Coefficient (MCC).
Our inclusion of these two metrics is due to the fact that many popular
cyber-attack (intrusion) datasets are highly imbalanced and
the standard metrics are misleading in the presence of data imbalance.
BA is defined as the arithmetic mean of sensitivity and specificity.
It is the better-known metric for imbalanced data. However,
when the class imbalance is extreme, which is common in publicly available intrusion datasets,
BA may prove unreliable and misleading just as the standard metrics, 
e.g., accuracy and F1-score \cite{Chicco2021-tv}.
In those cases metrics such as MCC are needed. MCC is
introduced by Matthews \cite{MATTHEWS1975442} and it is defined as 
\[ MCC = \frac{TP \times TN - FP \times FN}{\sqrt{(TP + FP) \times (TP + FN)  \times (FP + TN) \times (TN + FN))}}. \]

It ranges between -1 and +1, with -1 for perfect misclassification (TP = TN = 0) and 1 for perfect classification
(FP = FN = 0). MCC = 0 indicates random classification (TP $\times$ TN = FP $\times$ FN). 
It is undefined when a whole row or column of the confusion matrix is zero. 
However, it can be naturally extended for these cases: MCC := +1 when TP (or TN) is nonzero and all other
three entries are zero. MCC := -1 when FP (or FN) is nonzero and all other
three entries are zero. MCC := 0 for all the remaining cases where it is undefined. 
MCC is considered as the most accurate and reliable metric by many scientists, 
especially experts in bioinformatics, computational biology, brain informatics,  
and biomedical sciences \cite{Chicco2021-tv, Chicco2023-la}.
As we have not seen much adoption in cybersecurity, computer science, and general AI/ML,
we strive to make MCC more visible to these fields with our study.

\subsection{Related Work}
\label{relatedwork}
There has been a significant number of publications for interpretable and explainable ML
in cybersecurity in the last ten years \cite{surveyaicyber, moustafa2023explainable}.
The ML models that have been studied range from traditional ML (DTs, nearest neighbors methods, SVMs, ensembles etc.) 
to DL (feed-forward, convolutional, and recurrent DNNs, autoencoders etc.) \cite{surveyaicyber, moustafa2023explainable}.
The most used explanation methods are SHAPs \cite{shapley}, LIME \cite{lime}, and PDPs \cite{friedman2001greedy}. The applications of these methods
include botnet detection \cite{botnetdet}, malware detection \cite{malwaredet}, and 
intrusion detection \cite{mane2021explaining, SHARMA2024121751, KESHK2023119000}. 
In these applications, the most evaluated datasets include CIC-IDS2017 \cite{data2017},  NSL-KDD \cite{nslkdd}, CSE-CIC-IDS2018 \cite{Sharafaldin2018TowardGA}, CIC-DDoS2019 \cite{internetdataset}, TON\_IoT \cite{TON_IoT}, NF-TON-IoT V2 \cite{sarhan2022towards} CIDDS-001 \cite{CIDDS-001}, CIDDS-002 \cite{CIDDS-002} and 
UNSW-NB15 \cite{UNSWNB15}.

Some studies, such as \cite{KESHK2023119000, samarakoon20225g}, consider both binary and multi-class attack classification.
However, in real-world settings, intrusion detection systems are typically tasked to determine if there is an
attack or not while being online. They do not concern for determining the types of attacks since
intrusion monitoring is continuously performed in real-time \cite{Commercial_ids, BHARDWAJ2021100332} and the priority is
to detect the presence (or lack) of an attack. 
As a result, we focus on the binary classification of network traffic into attack and benign classes in our study.

Overall, we see that many published studies overlap in terms of
ML models, explanation methods and datasets. There is a common conduct in which many 
authors most often use SHAP and/or LIME to report feature scores 
for a trained ML model on one of the datasets we mention above and then conclude their study.
The problems we state in the introduction remain unattended and thus open.
This is what makes our study valuable. 
We empirically demonstrate the problems, study them, offer practical solutions, and highlight the key
research insights we gain.

\section{Experimental Setup and Datasets}
\label{experimentalsetup}
We choose two of the publicly available and real-world intrusion datasets for our experiments. One dataset is based on 5G network traffic \cite{5gdataset} and 
the other is based on traditional
Internet traffic \cite{internetdataset}. While there are other publicly available intrusion datasets as stated in Section \ref{relatedwork} and Section \ref{binaryclassificationmetrics}, these two datasets are
sufficient for our analysis: For the analysis and the experiments regarding DTs in Section \ref{classificationperfs}, there exist published studies conducting the same DTs experiments with the other datasets. We provide the corresponding references within that section. Our analysis in Sections \ref{binaryclassificationmetrics} and \ref{FeatureComputations} - \ref{Actionability} holds for the other datasets because the very underlying assumption of our analysis is the existence of aleatoric uncertainty. Such uncertainty exists (for every dataset) because of the probabilistic nature of data pre-processing and model training stages, the rampant approximations made throughout these two stages, and the approximations made during the generation of feature-based explanations. Lastly, our analysis in Section \ref{alignmentsection} is a theoretical one.

The first dataset is the real-world 5G flow traffic provided by Samarakoon et. al. \cite{samarakoon20225g, 5gdataset}. 
This dataset is generated using a real 5G testbed.
It contains several types of denial-of-service (DoS) attacks, such as ICMP, UDP, SYN and HTTP Floods and Slowrate DoS, and Port Scans such as SYN, TCP Connect and UDP Scan. 
We use the \textit{Encoded.csv} file of the dataset files that are published on the IEEE dataport \cite{5gdataset}. We take the  \texttt{Label} column as the true labels/classes. We set the benign entries to 0 and the malicious ones to 1. We then omit the columns \texttt{Unnamed:0}, \texttt{nan}, \texttt{Label}, \texttt{Attack Type}, \texttt{Attack Tool} 
and use the remaining features. 
The dataset consists of 1,215,890 flow entries. 
There are 477,737 benign flow entries out of 1,215,890, which is about 39.3\% of the whole dataset.
There are 96 fields in a flow entry and we use 91 of them as features for training.
Further details on the dataset can be found in the authors' paper \cite{samarakoon20225g} and on the IEEE dataport \cite{5gdataset}.

The second DoS dataset we use is the UDBLag part of the CIC-DDoS2019 dataset \cite{internetdataset}. 
The entire dataset, including the UDBLag part, is severely imbalanced in favor of attack flow entries. 
The UDPLag dataset has only 4016 benign flows out of a total of 674463 flows. 
That is, only about 0.6\% of the flows is benign and the rest is attack. 
This dataset has 88 fields and 77 of them are used for training.

We perform Python-based evaluations where we test each case 10$\times$ and report the mean scores.
Evaluations are performed on an Apple M2 Max system with MacOS Sonoma 14.4.1. 
The system has 12 CPUs, 38 GPUs, and 32 GB of memory.
The versions of Python and Scikit-Learn \cite{sklearn} that we use are 3.11.7 and 1.4.2, respectively.
The versions of Tensorflow \cite{tensorflow} and Keras \cite{keras} are 2.16.1 and 3.2.1, respectively.
We also use the SHAP python package (https://shap.readthedocs.io/en/latest/) whose version is 0.45.1. 

\section{Evaluation and Analysis}
\label{eval}
We now present our experimental evaluation and analysis.
Our analysis is eight-fold: 
\begin{enumerate}
\item evaluating the classification performance of ML models (Section \ref{classificationperfs}),
\item assessment of binary classification metrics (Section \ref{binaryclassificationmetrics}),
\item understanding the alignment among models and explanations (Section \ref{alignmentsection}),
\item obtaining feature importances intrinsically and through external methods \\
(Section \ref{FeatureComputations}), 
\item introducing cross-explanations (Section \ref{CrossExplanations}), 
\item studying strong feature correlations (Section \ref{FeatureCorrelations}), 
\item exploring the impact of hyper-parameters and optimization methods \\
(Section \ref{LearningProcess}), and
\item discussion of the utility of the explanations (Section \ref{Actionability}).
\end{enumerate}

\begin{table*}[ht]
\centering
\caption{Classification Scores of DTs, Ridge and DNNs for the 5G and UDBLag datasets}
\begin{tabular}{|l|l|l|l|l|l|l|} 
\hline
Classifier & \multicolumn{2}{|c|}{DT} & \multicolumn{2}{|c|}{Ridge} & \multicolumn{2}{|c|}{DNN}
\\ \hline
Metric  & 5G & UDBLag & 5G & UDBLag &  5G & UDBLag \\ \hline
 Accuracy & 0.9995920 &  0.9999604 & 0.9912623 & 0.9993575
 &  0.9996107 & 0.9998121 \\  \hline 
 BA & 0.9995873 &  0.9983037 & 0.9900997 & 0.9596470  
 & 0.9996554 & 0.9958653 \\ \hline
 F1 &  0.9997179 & 0.9999801  & 0.9900906 & 0.9995228 
 & 0.9996805 & 0.9999055 \\ \hline
 Precision & 0.9996094 &  0.9999801 &  0.9955576 & 0.9998309 
 & 0.9999099 & 0.9999502 \\ \hline
 Recall & 0.9996636 &  0.9999801 & 0.9928166 & 0.9996769 
 & 0.9994512 & 0.9998607 \\ \hline
 MCC & 0.9991453  & 0.9966074 & 0.9816909 & 0.9440313 
 & 0.9991824 & 0.9845968 \\ \hline
\end{tabular}
\label{tab:all_scores}
\end{table*}
\subsection{Classification Performance}
\label{classificationperfs}
We now present the binary classification performances of Ridge Classifiers and
DTs that are implemented in Scikit-Learn library \cite{sklearn}, and 
a Keras-based DNN \cite{keras}. The DNN has two fully connected hidden layers, 
each with ten nodes and \texttt{Relu} activation. The output layer is a node with \texttt{Sigmoid} activation.
Binary cross entropy is selected as the loss function.
We consciously choose a DNN that is as simple as possible. As we report next, this simple DNN performs very well. In addition, we use this two-layered DNN architecture for all relevant evaluations in our study. 
Except for the evaluations presented in Subsection \ref{LearningProcess},
we use the default Keras optimizer \texttt{RMSprop} and we set the batch size to 256 and the number of the epochs to 5.

Table \ref{tab:all_scores} shows the average classification scores of DTs, Ridge classifiers, and DNNs. 
DTs and Ridge classifiers are trained with default values.
We see that for both 5G and UDBLag datasets, DTs perform extremely well and
achieve scores that are all $>=$ 0.996. In particular,
for the 5G dataset, for all metrics, DTs achieve scores that are $>=$ 0.999. 
We report the results in high precision which is valuable in real-world deployments.
As for Ridge classifiers, while not as good as those of DTs, 
they achieve scores $>=$ 0.98 for the 5G dataset and $>=$
0.944 for the UDBLag dataset. These scores show that 
simple linear classifiers perform well for both datasets.  
As for DNNs, we see that for the 5G dataset, for all metrics, 
DNN achieves scores that are $>=$ 0.999 - the same with DTs. For the UDBLag dataset, all scores $>=$ 0.99, except MCC which is $>=$ 0.985.

Many intrusion detection studies have found that DTs perform extremely well
with the available intrusion datasets. To name a few, the authors of the 5G dataset  Samarakoon et. al. \cite{samarakoon20225g}
report that DTs achieve accuracy, precision, recall, and F1 scores that are all $>= 0.998$ (Table X (10) in \cite{samarakoon20225g})
for the dataset. Mahbooba et. al. \cite{Mahbooba2021in} state that DTs achieve perfect precision, recall, and F1 scores (Figure 4 in \cite{Mahbooba2021in}) for NSL-KDD \cite{nslkdd}. Gaitan-Cardenas, Abdelsalam, and Roy \cite{dt-cidds-eval}
report that DTs obtain accuracy, precision, recall, and F1 scores that are all $>= 0.999$ for the CIDDS-001 \cite{CIDDS-001} and NF-TON-IoT V2 \cite{sarhan2022towards} datasets. They also
report that DTs achieve perfect scores of 1 for all metrics for the CIDDS-002 dataset \cite{CIDDS-002}
(Table II in \cite{dt-cidds-eval}).
For the UNSW-NB15, the study \cite{UNSW-NB15Analysis} (Table 7) reports that 
DTs achieve accuracy, precision, recall, and F1 scores that are all
$>= 0.99$.
Neto et. al. \cite{CICIoV2024ids} report that AdaBoost based on DTs obtains perfect scores for
accuracy, precision, recall, and F1 for the CICIoV2024 (Figure 6a with decimal encoding of the data). 
For the CIC-IDS2018, Songma, Sathuphan and Pamutha \cite{CIC-IDS-2018study} report that
DTs' accuracy, precision, recall, F1, BA, and MCC are 0.99, 0.96, 0.97, 0.96, 0.97, and 0.999, respectively. 
They further provide experiments with improved DT implementations having higher scores. 
Rosay et. al. \cite{CIC-IDS2017study} (Table 2) report DT scores $>= 0.99$ for accuracy, precision, recall, and MCC for the CIC-IDS2017.

\newcounter{mycounter}
\newcounter{myadvise}
\setcounter{myadvise}{1}
\setcounter{mycounter}{1}
\colorlet{shadecolor}{red!10}
\begin{shaded*}
\textbf{Main Result \themycounter:}
\stepcounter{mycounter}
\textit{Therefore, 
in light of our results and many published studies,
we stress that the existing intrusion datasets
do not require the usage of complex high-cost black-box models such as DNNs 
since the problem of binary attack classification for these datasets is
completely solvable by DTs, which are highly interpretable and low-cost ML models.
}
\end{shaded*}

\colorlet{shadecolor}{green!10}
\begin{shaded*}
\textbf{Guide \themyadvise:}
\stepcounter{myadvise}
DTs, their variants (e.g., gradient boosted trees) and their ensembles (e.g., random forests) 
must be the starter ML models for intrusion detection.
\end{shaded*}

\subsection{Proper Binary Classification Metrics}
\label{binaryclassificationmetrics}
The usage of proper binary classification metrics is a key
component of accurate and reliable evaluation of ML classifiers. 
To showcase
this, we train a DNN with the features obtained by SHAPs as in 
Figure \ref{fig:dnn_shap_inter_2_9876} for the UDBLag dataset. 
From the figure, we see that
the top three features are \texttt{ACK Flag Count}, \texttt{URG Flag Count}, and \texttt{Min Packet Length}.
Table \ref{tab:dnn_proper_res} shows the DNN's classification scores with these features. We see that even though all
standard metrics, i.e., accuracy, F1, precision, and recall, are $>= 0.998$, BA and MCC
are 0.8967 and 0.8672, respectively.   
The standard metric scores are very misleading. 
Only after computing BA and MCC, 
the true classification performance is known. 
To see the true picture, we check the confusion matrix which is
$\begin{bmatrix}
504 &  131 \\
27 & 100508
\end{bmatrix}$. From the matrix, we see that the DNN's performance is actually poor.
For instance, the false positive rate is $\frac{131}{504+131} = 0.206$, 
which is unacceptable and unsustainable for real-world intrusion detection systems. 
Such a system would raise a false alarm once for every five predictions made on average.

\begin{table}[ht]
\centering
\caption{DNN classification scores with the features \texttt{ACK Flag Count}, 
\texttt{URG Flag Count}, and \texttt{Min Packet Length} for the UDBLag dataset.}
\begin{tabular}{|l|c|l|c|} 
\hline
 Metric & Score & Metric & Score\\ \hline
 Accuracy &  0.9984382722150835 & Precision &  0.9986983177495802 \\ \hline
 BA &  0.8967161121073125 & Recall &  0.9997314368130502 \\ \hline
 F1 & 0.9992146102379035 & MCC & 0.8672112869253433 \\ \hline
\end{tabular}
\label{tab:dnn_proper_res}
\end{table}
\begin{table}[ht]
\centering
\caption{Binary Class Distribution of Intrusion Detection Datasets}
\begin{tabular}{|l|l|l|c|c|}
\hline
Dataset & Total Entries & Attack Entries & Attack - Benign & Severity\\ \hline
5G \cite{5gdataset} & 1215890 & 738153 & 61\% - 39\% &  Mild  \\ \hline
UDBLag \cite{internetdataset} & 674463 & 670447 & 99\% - 1\% & Extreme \\ \hline
CIC-DDoS2019 \cite{internetdataset} & 50063112 & 50006249 & 99.9\% - 0.1\% & Extreme \\ \hline
CIC-IDS2018 \cite{Sharafaldin2018TowardGA} & 10974408 & 1865649 & 17\% - 83\% & Moderate  \\ \hline
CIC-IDS2017 \cite{data2017} & 2810677 & 2359087 & 84\% - 16\%  & Moderate \\ \hline
NSL-KDD  \cite{nslkdd} & 311027 & 250436 & 81\% - 19\% & Moderate  \\ \hline
NF-TON-IoT-V2 \cite{sarhan2022towards} & 16940496 & 10841027 & 64\% - 36\% & Mild \\ \hline
CIDDS-001 \cite{CIDDS-001} & 172838 & 107343 & 62\% - 38\% & Mild  \\ \hline
CIDDS-002 \cite{CIDDS-002} & 1048576 & 699051 & 67\% - 33\%  & Mild  \\ \hline
UNSW-NB15 \cite{UNSWNB15} & 2540044 & 321283 & 13\% - 87\% & Severe  \\ \hline
CICIoV2024 \cite{NETO2024101209} & 1408219 & 184482 & 13\% - 87\% & Severe  \\ \hline
CICEV2023 \cite{CICEV2023} & 63284 & 58000 & 92\% - 8\% & Severe  \\ \hline
\end{tabular}
\label{tab:class_imbalance}
\end{table}

For the next step, we collect statistical information regarding the binary class distribution for the publicly available
intrusion detection datasets. This is to see if there is any class imbalance in the datasets.
Table \ref{tab:class_imbalance} shows the binary class details for twelve datasets.
In particular, it shows the proportions of attack and benign entries. We see that all datasets
have binary class imbalance. Among the datasets, two datasets have 
a majority-class proportion that is $>=$ 99\%, another six datasets have 
a majority-class proportion that is $>$ 80\% and $<$ 99\%. 

To characterize the datasets qualitatively, we refine the definition of
the degree of imbalance provided by \cite{Imbalancedata}. If in a dataset, the majority class 
has a proportion that is $>=$ 60\% and $<$ 75\%, then the degree of imbalance is defined as \textit{mild}. 
If the majority class's proportion is 
$>=$ 75\% and $<$ 85\%, then the degree of imbalance is said to be \textit{moderate}.
If the majority class's proportion is 
$>=$ 85\% and $<$ 99\%, then the degree of imbalance is called \textit{severe}. 
Finally, if it is $>=$ 99\%, the degree of imbalance is called \textit{extreme}. 
According to this qualitative characterization, we see that out of twelve datasets, 
eight datasets have a degree of imbalance that is at least \textit{moderate}. 
Two datasets have \textit{extreme} imbalance while another three have \textit{severe} imbalance.
Overall, we conclude that a majority of the existing intrusion detection datasets 
have a significant level of binary class imbalance. 

\colorlet{shadecolor}{red!10}
\begin{shaded*}
\textbf{Main Result \themycounter:}
\stepcounter{mycounter}
\textit{Given that the existing intrusion datasets are typically highly imbalanced, standard classification metrics might be very misleading and overestimating the true performance. One particular reason for this is that even when all the standard scores are bigger that 0.99, the true classification performance might be much worse.
}
\end{shaded*}

\colorlet{shadecolor}{green!10}
\begin{shaded*}
\textbf{Guide \themyadvise:}
\stepcounter{myadvise}
Matthews Correlation Coefficient (MCC) and balanced accuracy (BA) are a must for reliable and accurate evaluation of binary classification performance.
\end{shaded*}

\begin{figure*}[ht]
    \centering
    \begin{subfigure}[t]{0.45\textwidth}
    \includegraphics[width=1\linewidth]{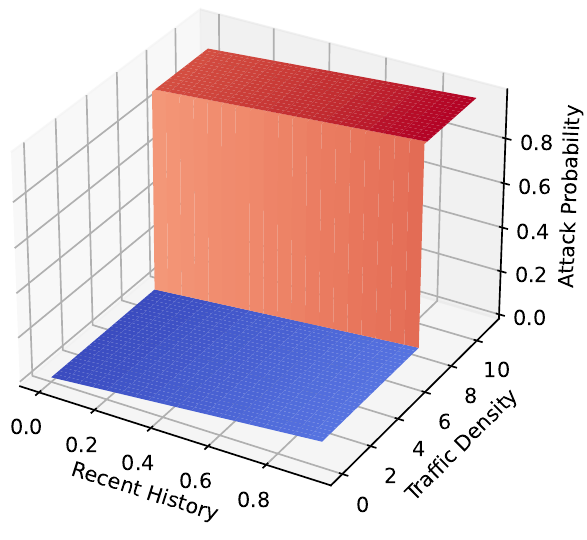}
    \caption{$M_{step} = c_1 (T > threshold) + c_2 H\\
    \qquad M_{step} = 0.9 \times (T > 7) + 0.1 \times H$}
    \label{fig:dnn_definition_m1}
    \end{subfigure}\hfill
    \begin{subfigure}[t]{0.45\textwidth}
    \includegraphics[width=1\linewidth]{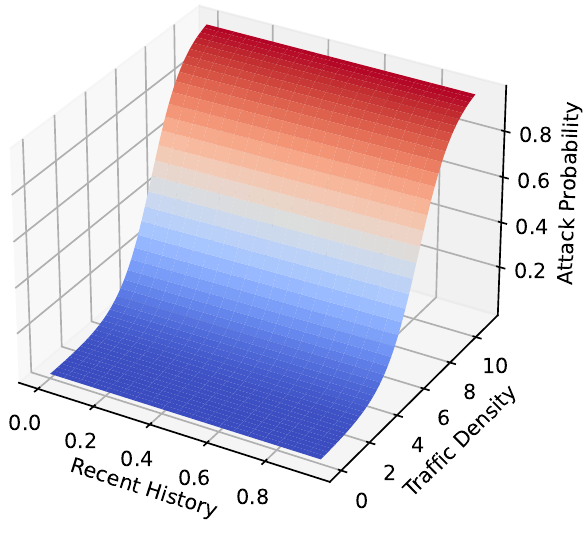}
    \caption{$M_{smooth} = \sigma(c_1 (T - threshold) + c_2  H)$\\
    \qquad$M_{smooth} = \sigma(0.9 \times (T - 7) + 0.1 \times H)$}
    \label{fig:dnn_definition_m2}
    \end{subfigure}\hfill
    \caption{The plots of models $M_{step}$ and $M_{smooth}$ where $c_1 = 0.9$, $c_2 = 0.1$, and $threshold = 7$.
    The plot is inspired by Chapter 33 - Section 33.1.2 of \cite{pml2Book}.}
    \label{fig:dnn_explanation_definition}
\end{figure*}

\subsection{Alignment Among Models and Explanations}
\label{alignmentsection}
Considering the state of the art in interpretable and explainable ML for intrusion detection,
there is a need for more detailed studies that are conducted with both experimental and mathematical rigor. 
In this section, we use a simplified example where we show that if we are not careful enough, 
the accurate alignment between the specification of an explanation, 
the computation of that explanation, and the target model may be compromised. 
The model may become misaligned with the intended computation of the explanation.  
Ultimately, this misalignment may lead to incorrect explanations.

As a starter, we assume that we use two different 
ML models for predicting the probability of a cyber attack. 
Let these models be the following simple, linear and interpretable mathematical models:
\[ M_{step} = c_1 \times (T > threshold) + c_2 \times H \]
\[ M_{smooth} = \sigma(c_1 \times (T - threshold) + c_2 \times H) \]
where the feature traffic density $T$ represents the average traffic density per day. We assume $T$ be between 0 and 10.  
Moreover, the feature recent history $H$ represents the number of weeks that have seen at least 
one attack over the last ten weeks. That is,
$0<= H =< 1$ and $H = 1$ means that there has been at least one attack per each of the last ten weeks.
$H = 0.5$ corresponds to the scenario where there was at least one attack in 
five different weeks over the last ten weeks.
We choose these models such that they are very similar and inherently interpretable. 
The only differences between the models are the usage of a 
\textit{sigmoid function} $\sigma(x) = \frac{1}{1 + e^{-x}}$  in $M_{smooth}$ and
the application of a threshold ($threshold$). 
Proceeding further, let us assume that both models are trained and
their trained parameters turn out to be the same, where
$c_1 = 0.9$, $c_2 = 0.1$, and $threshold = 7$. With these parameter values,
Figure \ref{fig:dnn_explanation_definition}
visualizes the models and the colorized gradients with respect to the features 
traffic density $T$ and recent history $H$.

We want to identify the most important feature in predicting the probability of an attack.
If we define the importance of a feature as to what extent the feature dominates 
other features in determining a model's prediction, then both $M_{step}$ and $M_{smooth}$
agree on the feature traffic density $T$ due to its coefficient (0.9 vs 0.1).
In contrast, if the importance is defined by the magnitude of the derivatives (gradients) of the prediction 
with respect to the inputs, that is, if it is based on local geometry and curvature, then for $M_{smooth}$,
the most important feature is still the traffic density $T$.
However, for $M_{step}$, the most important feature becomes the recent history $H$. 
This is because the local gradient/curvature of $H$ is 0.1 everywhere, 
while the local gradient of $T$ is zero almost everywhere, except where the threshold is located. 
Without carefully analyzing how a change in the definition of the feature importance affects
a model explanation, we could have incorrectly concluded that the most important feature for 
$M_{step}$ would remain the same just because it was very similar to $M_{smooth}$.
\colorlet{shadecolor}{green!10}
\begin{shaded*}
\textbf{Guide \themyadvise:}
\stepcounter{myadvise}
\textit{Even for the simplest mathematical models,
rigorous analysis is required when computing a model explanation. 
Conducting such analysis helps prevent unaccounted disagreements between the model and its intended
computation of explanations.
}
\end{shaded*}

\subsection{Feature Importance Computations}
\label{FeatureComputations}
\begin{figure*}[ht]
    \centering
    \begin{subfigure}[t]{0.31\textwidth}
    \includegraphics[width=\linewidth]{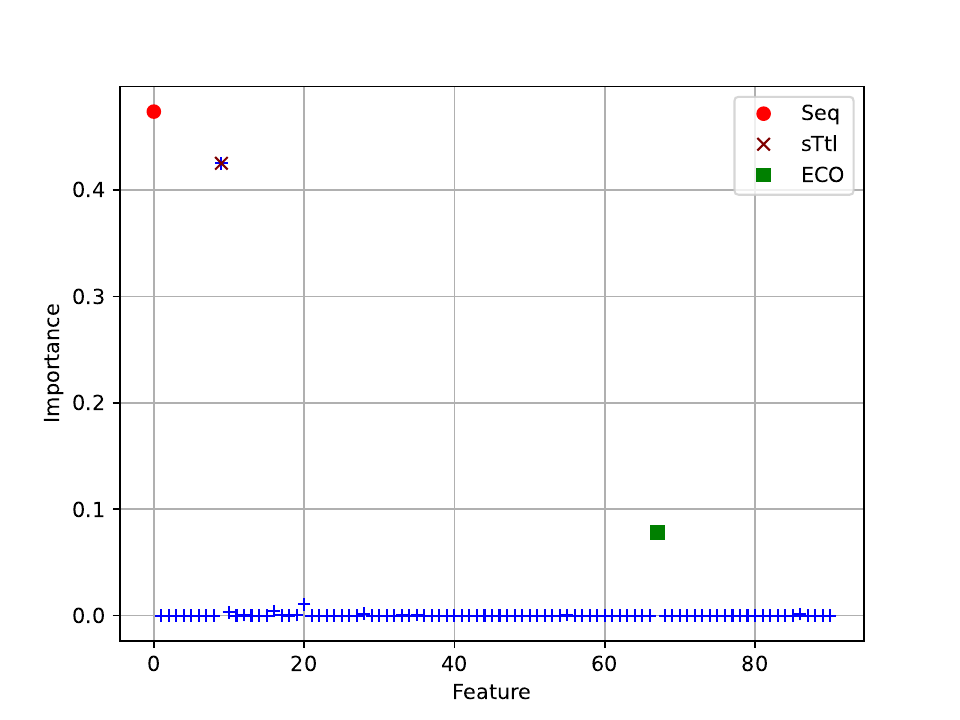}
    \caption{FIs for DT.}
    \label{fig:dt_self_feat_importance_5g}
    \end{subfigure}\hfill
    \centering
    \begin{subfigure}[t]{0.3\textwidth}
    \includegraphics[width=\linewidth]{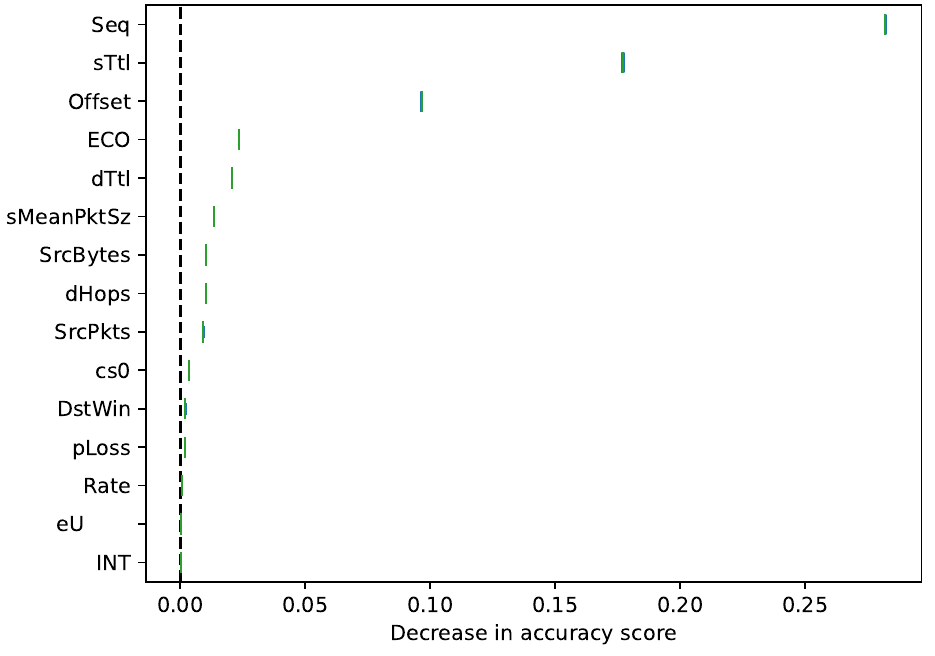}
    \caption{PIs for DT.}
    \label{fig:dt_per_importance_5g_2}
    \end{subfigure}\hfill
    \centering
    \begin{subfigure}[t]{0.35\textwidth}
    \includegraphics[width=\linewidth]{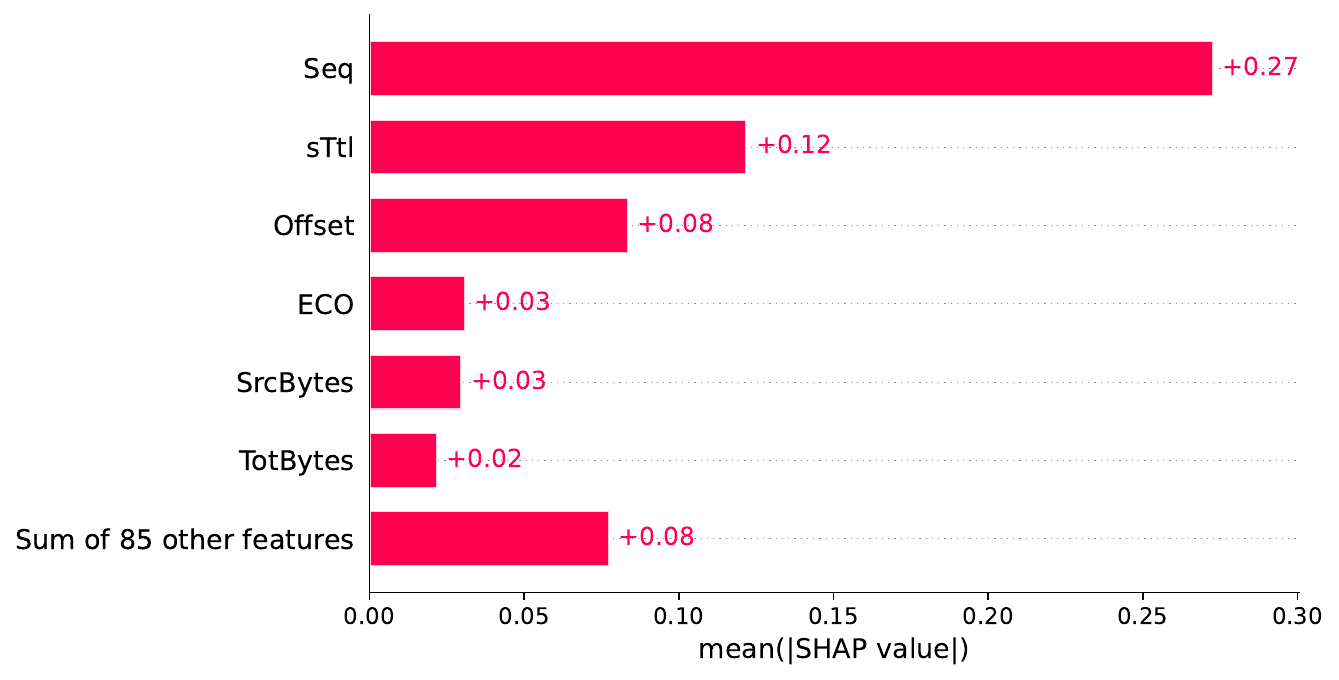}  
    \caption{SHAPs for DT.}
    \label{fig:dt_shapley_5g}
    \end{subfigure}
    \caption{The top features based on the FIs, PIs, and SHAPs for DTs for the 5G dataset.}
\label{fig:5g_unified_feature_scores}
\end{figure*}

In this section, 
we evaluate and discuss the Feature Coefficients (FCs) of Ridge, Feature Importances (FIs) of DTs, and
Permutation feature Importances (PIs) and SHapley Additive exPlanations (SHAPs) for Ridge, DTs, and DNNs. 
Figures \ref{fig:dt_self_feat_importance_5g}, \ref{fig:dt_per_importance_5g_2} and \ref{fig:dt_shapley_5g} 
show the top feature scores computed intrinsically by DTs themselves, i.e, FIs, by PIs,
and by SHAPs, respectively  
when DTs are trained on the 5G dataset. Focusing only on the top three features,
we see that all three methods agree on the top two features, 
which are \texttt{Seq} and \texttt{sTtl}. These two top features 
do not seem to change across different DTs or random seeds in our evaluations.

\begin{figure*}[!ht]
    \centering
    \begin{subfigure}[t]{0.31\textwidth}
    \includegraphics[width=1\linewidth]{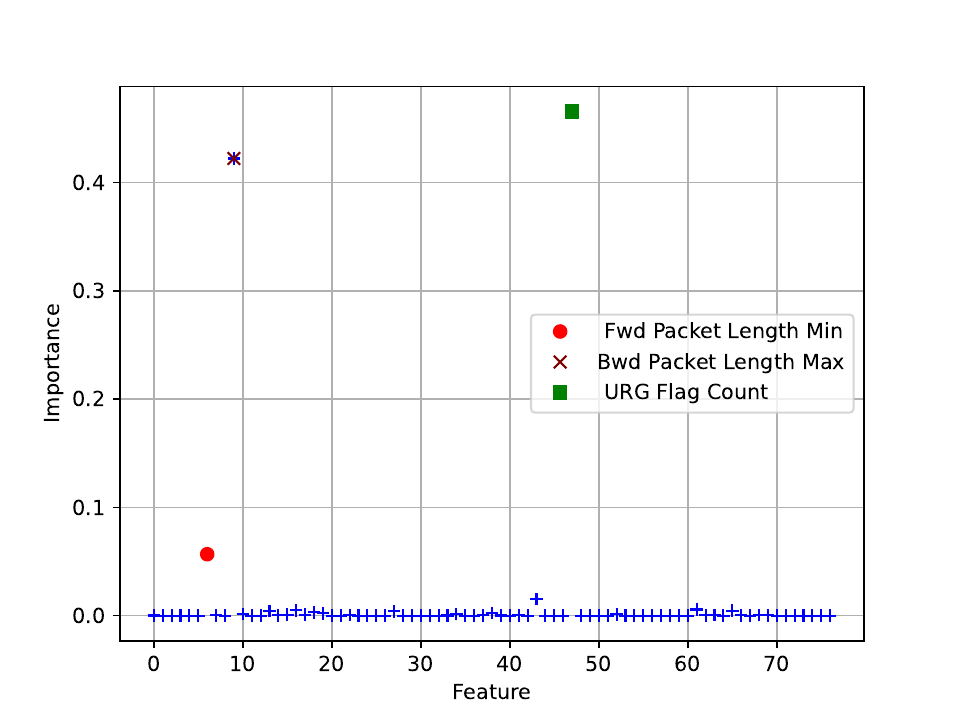}
    \caption{FIs for DTs with random seed 1234.}
    \label{fig:dt_self_feat_importance_inter}
    \end{subfigure}\hfill
    \centering
    \begin{subfigure}[t]{0.31\textwidth}
    \includegraphics[width=1\linewidth]{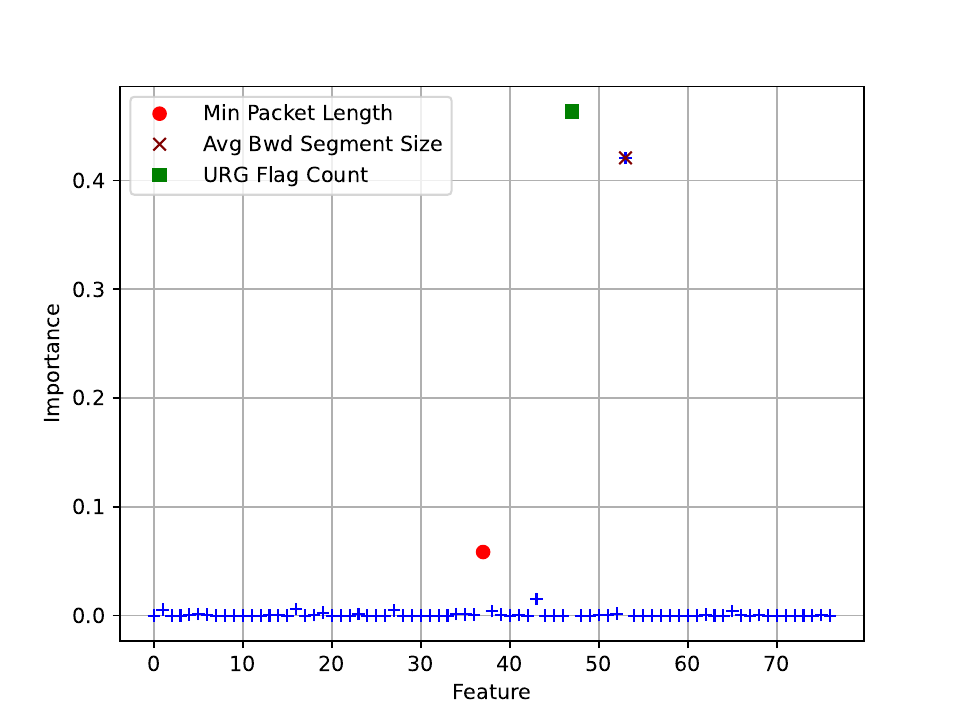}
    \caption{FIs for DTs with random seed 5678.}
    \label{fig:dt_self_feat_importance_inter2}
    \end{subfigure}\hfill
    \centering
    \begin{subfigure}[t]{0.31\textwidth}
    \includegraphics[width=1\linewidth]{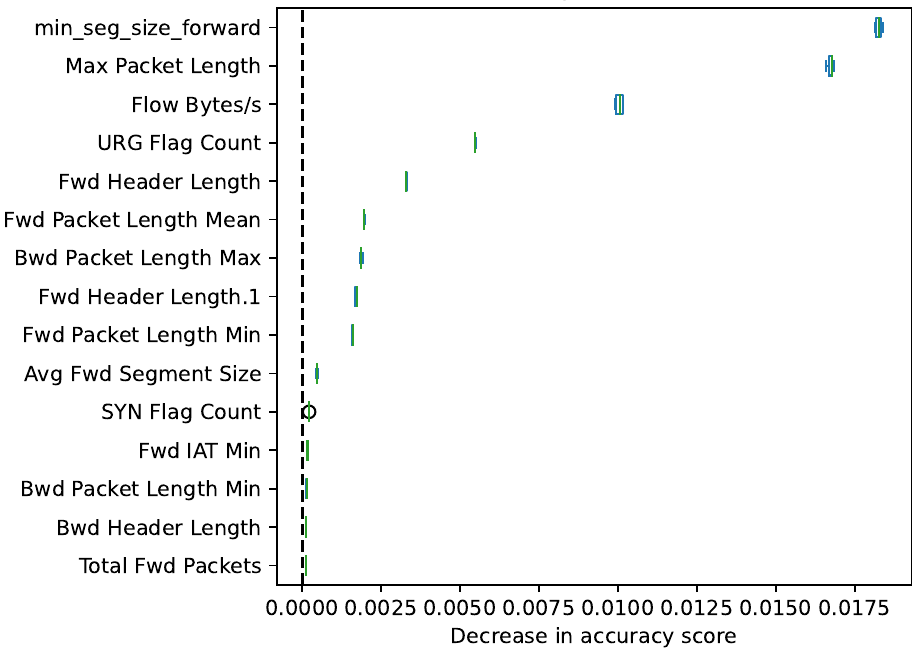}
    \caption{PIs for DT with random seed 1234.}
    \label{fig:dt_per_importance_inter}
    \end{subfigure}\hfill
    \centering
    \begin{subfigure}[t]{0.31\textwidth}
    \includegraphics[width=1\linewidth]{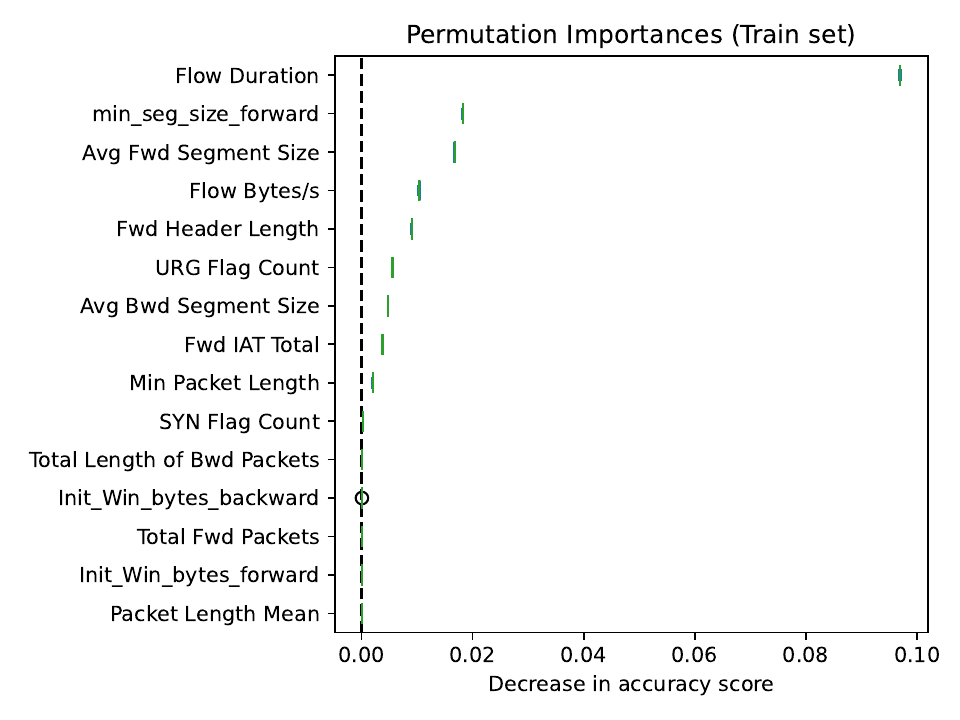}
    \caption{PIs for DT with random seed 6789.}
    \label{fig:dt_per_importance_inter_2}
    \end{subfigure}\hfill
    \centering
    \begin{subfigure}[t]{0.33\textwidth}
    \includegraphics[width=1\linewidth]{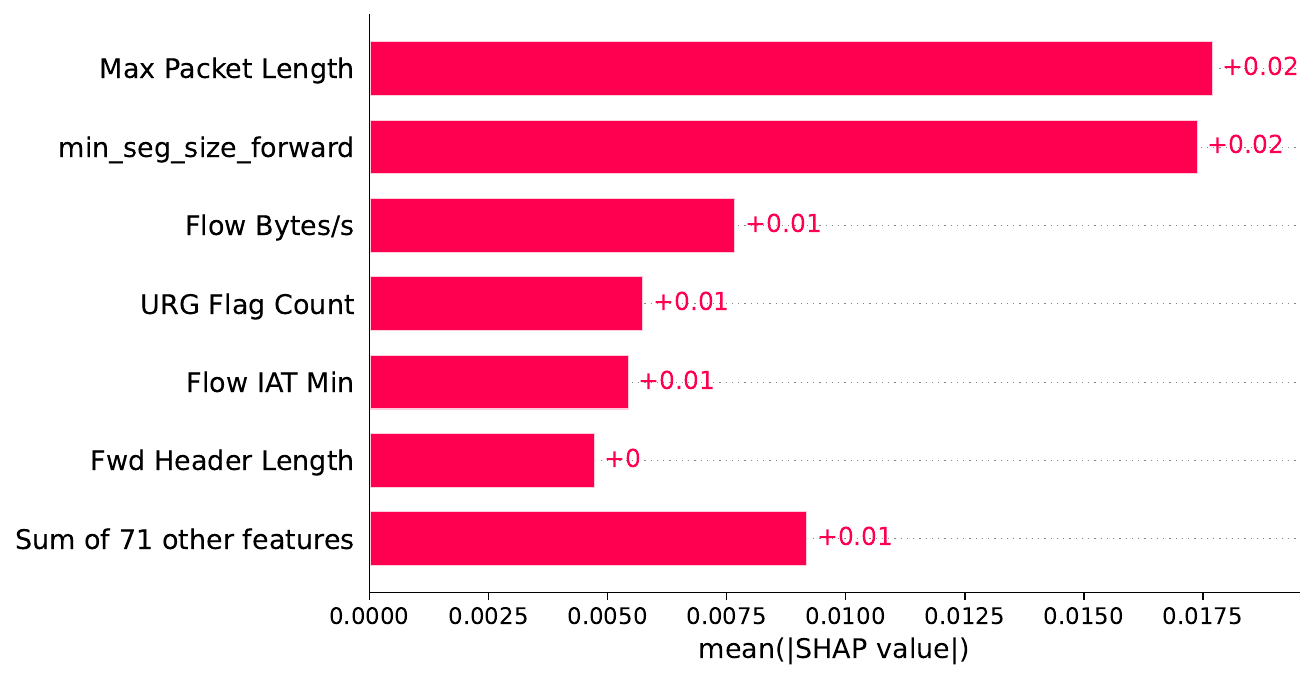}  
    \caption{SHAPs for DT with random seed 1234.}
    \label{fig:dt_shapley_udb_1234}
    \end{subfigure}\hfill
    \centering
    \begin{subfigure}[t]{0.33\textwidth}
    \includegraphics[width=1\linewidth]{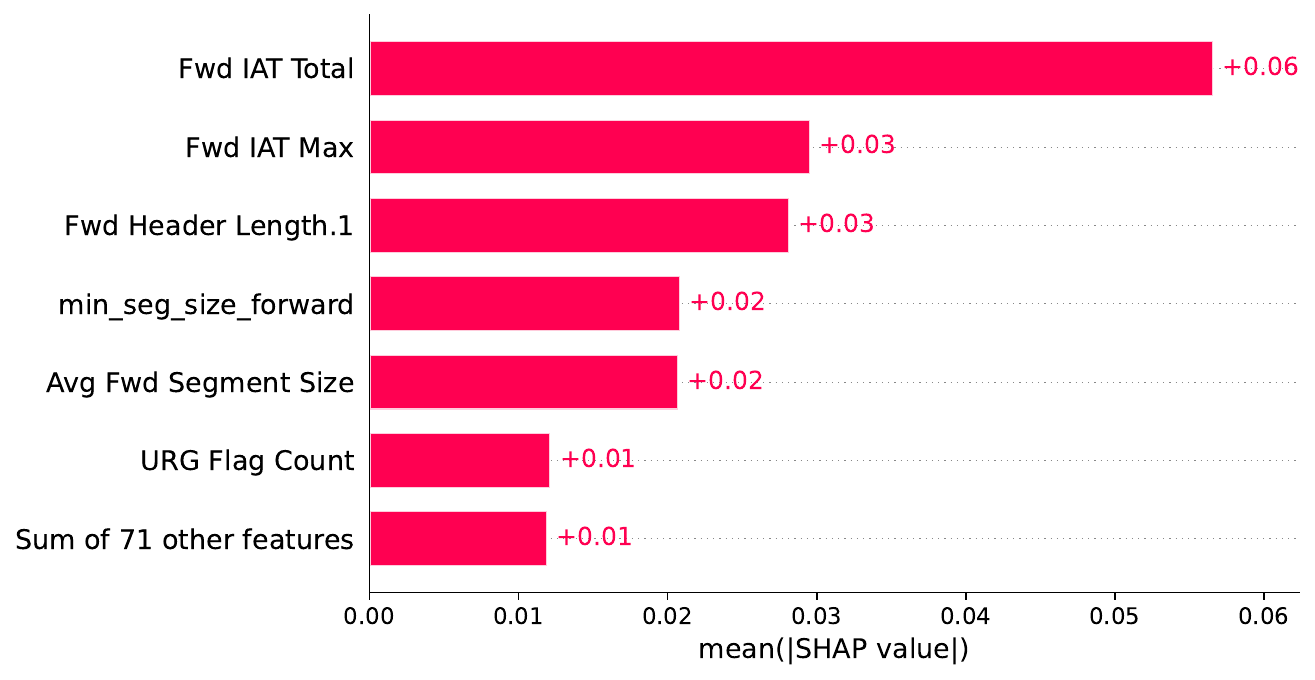} 
    \caption{SHAPs for DT with random seed 5678.}
    \label{fig:dt_shapley_udb_5678}
    \end{subfigure}
\caption{The top features based on the FIs, PIs, and SHAPs for 
DTs for the UDBLag dataset with two different random seeds.}
\label{fig:dt_udblag_unified_feature_scores}
\end{figure*}

Figure \ref{fig:dt_udblag_unified_feature_scores} 
shows the top features computed by FIs, by PIs,
and by SHAPs  
when DTs are trained on the UDBLag dataset.
We include two exemplary results for each type of scores obtained.
This is because
when different random seeds are used, the output of each method
differs. We see that the resulting top three features
differ for the same explanation method and across different explanation methods.
For instance, for the FIs, only the third top feature is the same, i.e., \texttt{URG Flag Count}.
Across all six results in Figure \ref{fig:dt_udblag_unified_feature_scores}, 
the very top feature is different than others. 
Moreover, in contrast to the 5G dataset, 
the top two features are not the same across different settings.

\begin{figure*}[t]
    \centering
    \begin{subfigure}[t]{0.43\textwidth}
    \includegraphics[width=1\linewidth]{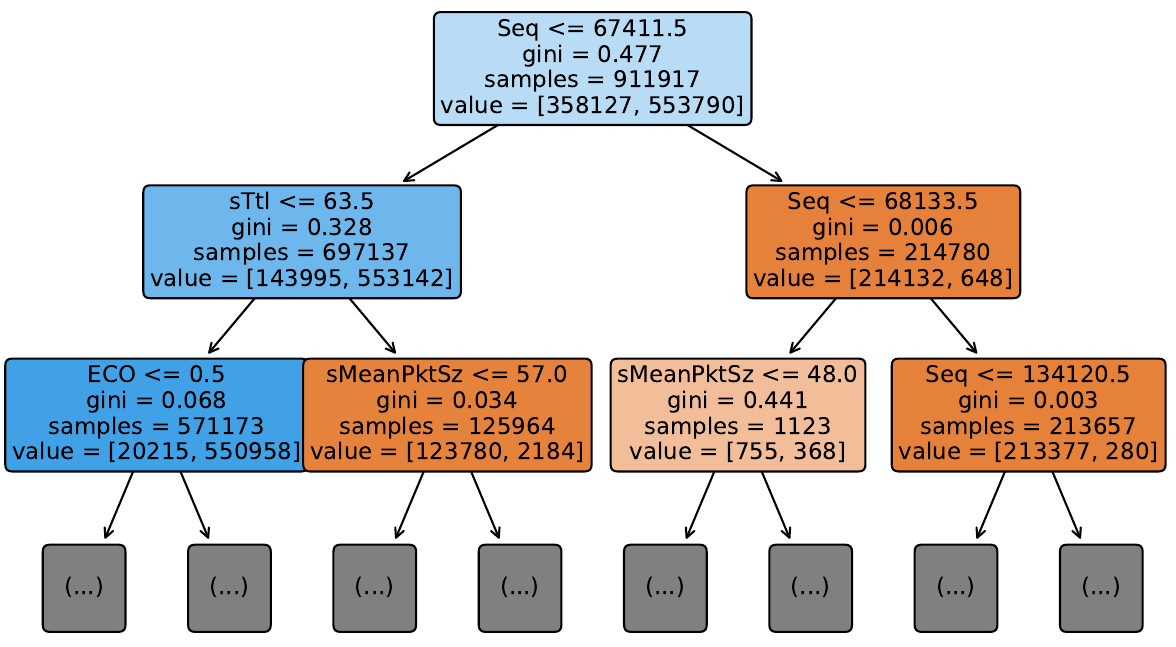}
    \caption{The resulting DT for the 5G dataset.}
    \label{fig:data_tree_5g}
    \end{subfigure}\hfill
    \centering
    \begin{subfigure}[t]{0.56\textwidth}
    \includegraphics[width=1\linewidth]{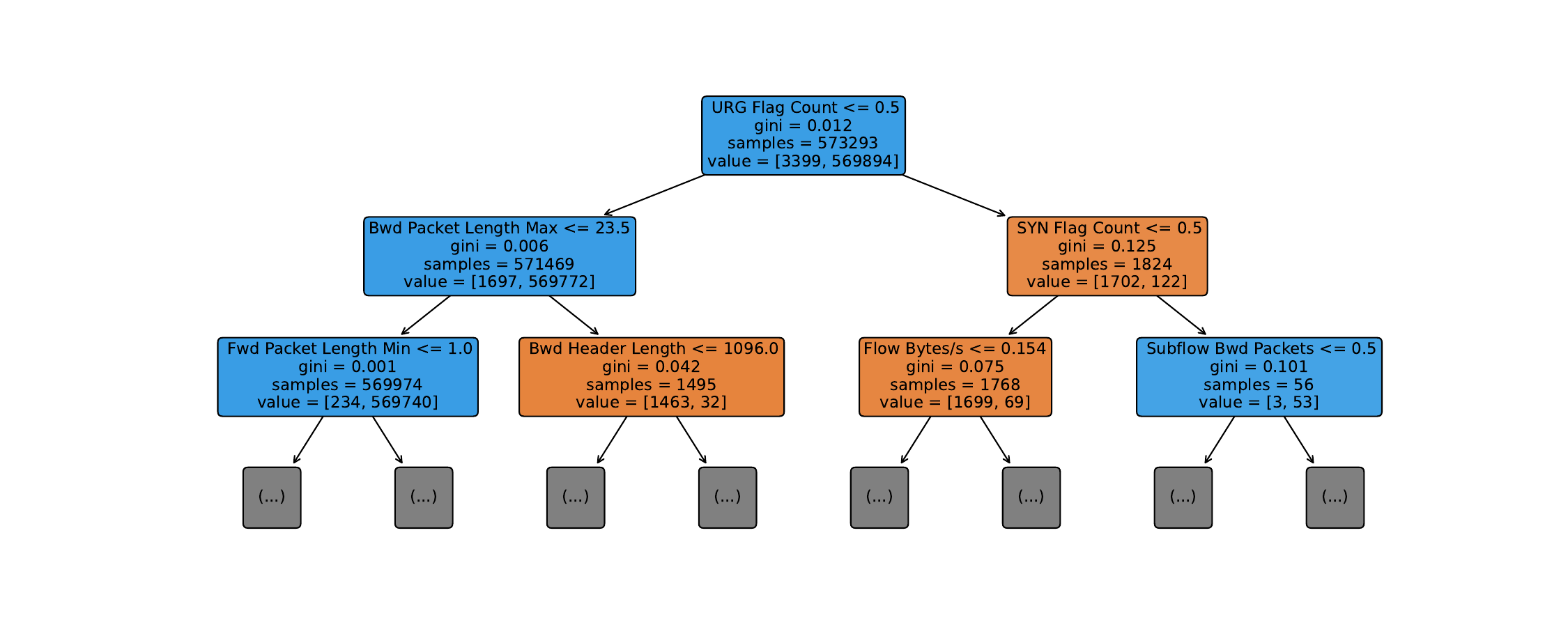}
    \caption{The resulting DT for the UDBLag dataset.}
    \label{fig:data_tree_inter}
    \end{subfigure}
 \caption{The resulting DTs for the two datasets.}
\label{fig:resulting_dts}
\end{figure*}

Figures \ref{fig:data_tree_5g} and \ref{fig:data_tree_inter}
show the resulting DTs for the 5G and the UDBLag dataset.
For ease of illustration, we just plot the top three levels in the trees.
We can clearly reason about how a DT classifies a flow instance (entry) by
following the path from the root to the leaf node and using boolean logic along the path.
That is, conjoining the node conditions along the path constructs the exact boolean expression 
that leads to the classification.
Here, we include just two examples of the resulting DTs for brevity. 
As with the most impactful features, 
the resulting DTs differ with different random seeds.

\begin{figure*}
    \centering
    \begin{subfigure}[t]{0.32\textwidth}
    \includegraphics[width=1\linewidth]{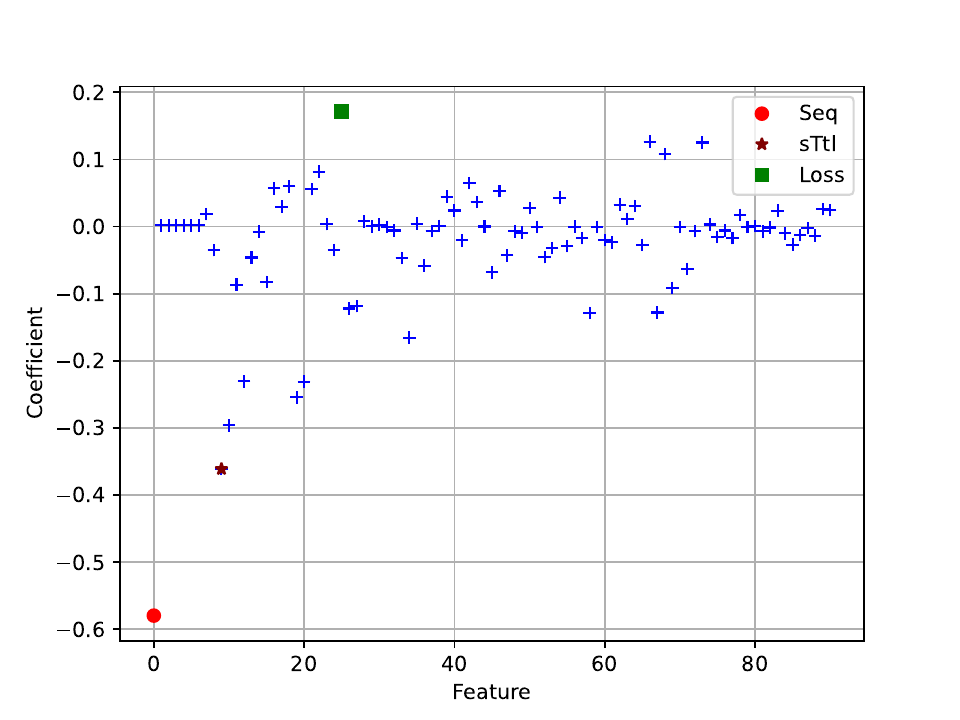}
    \caption{FCs of Ridge with the 5G dataset.}
    \label{fig:rg_feat_coefs_5g}
    \end{subfigure}\hfill
    \centering
    \begin{subfigure}[t]{0.32\textwidth}
    \includegraphics[width=1\linewidth]{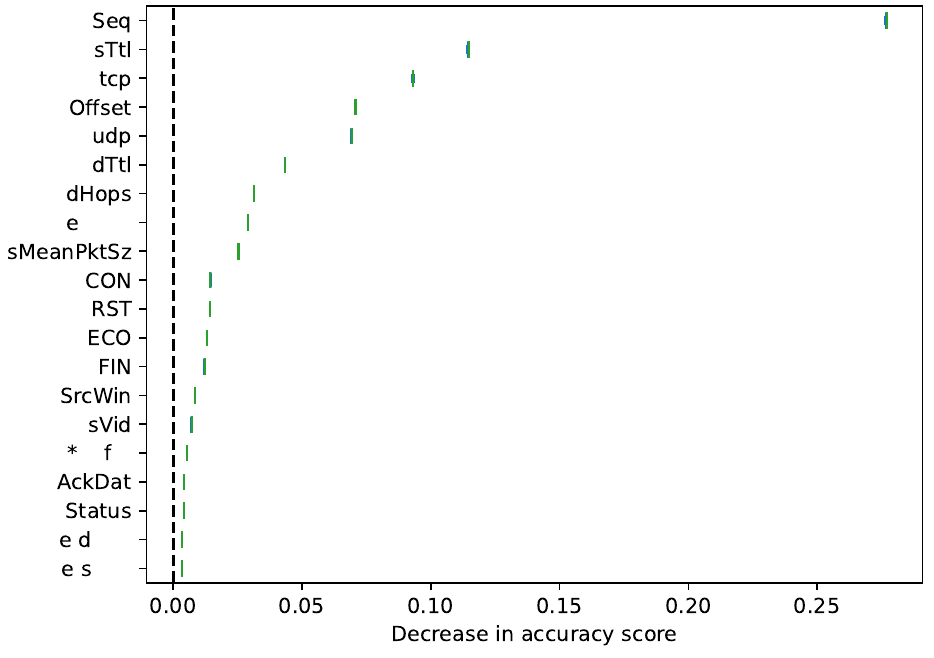}
    \caption{PIs for Ridge with the 5G dataset.}
    \label{fig:rg_feat_permutation_5g}
    \end{subfigure}\hfill
    \centering
    \begin{subfigure}[t]{0.32\textwidth}
    \includegraphics[width=1\linewidth]{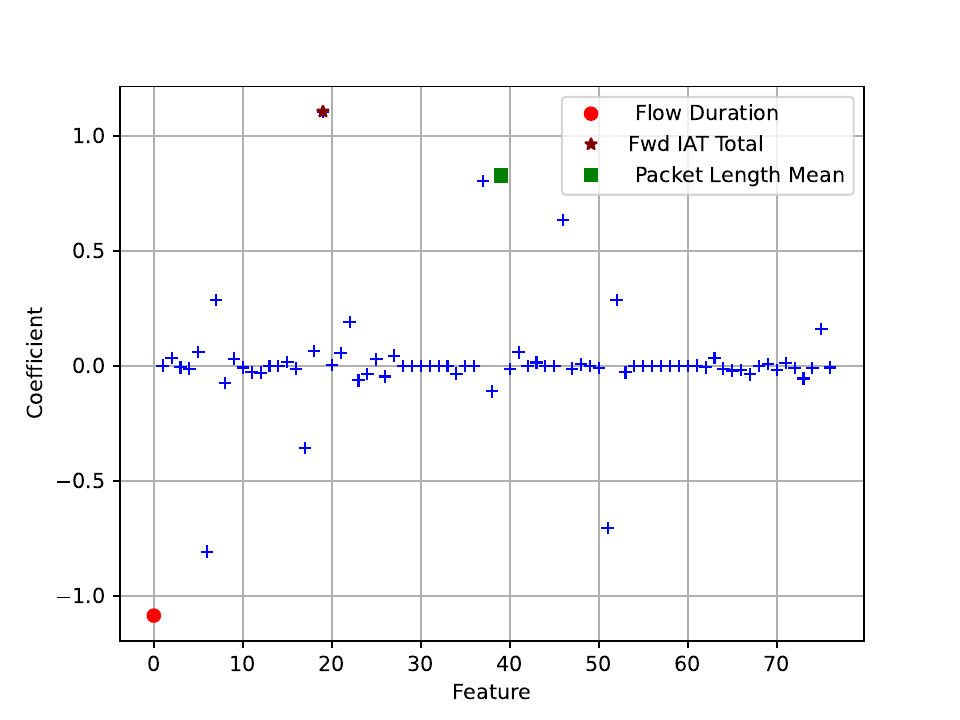}
    \caption{FCs of Ridge with the UDBLag dataset with random seed 1234.}
    \label{fig:rg_feat_coefs_inter}
    \end{subfigure}\hfill
    \centering
    \begin{subfigure}[t]{0.32\textwidth}
    \includegraphics[width=1\linewidth]{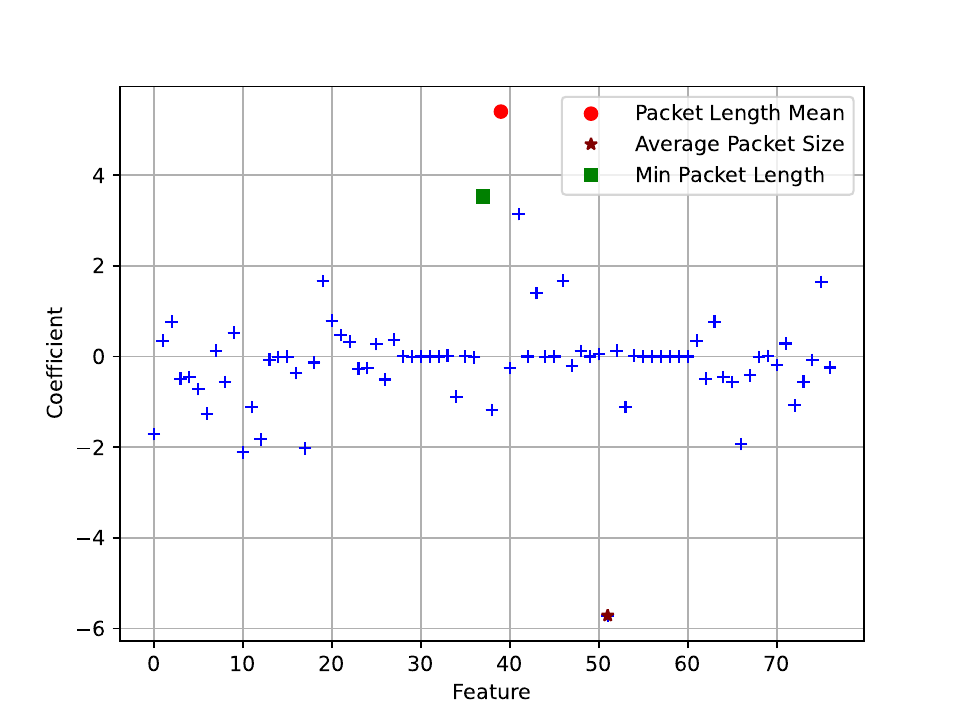}
    \caption{FCs of Ridge with the UDBLag dataset with random seed 5678.}
    \label{fig:rg_feat_coefs_inter2}
    \end{subfigure}\hfill
    \centering
    \begin{subfigure}[t]{0.32\textwidth}
    \includegraphics[width=1\linewidth]{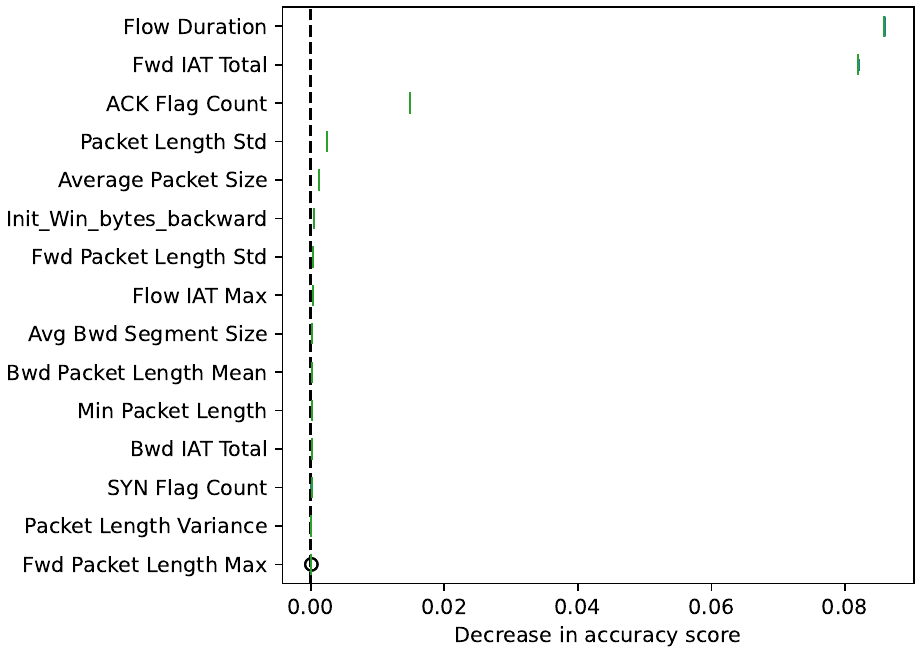}
    \caption{PIs for Ridge with the UDBLag dataset with random seed 1234.}
    \label{fig:rg_permutation_inter}
    \end{subfigure}\hfill
    \centering
    \begin{subfigure}[t]{0.32\textwidth}
    \includegraphics[width=1\linewidth]{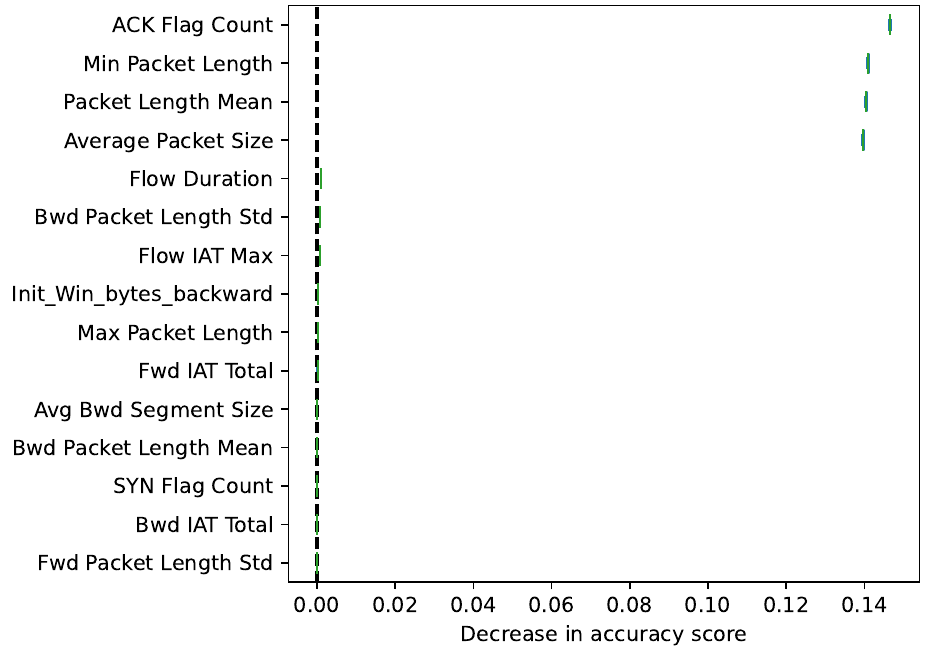}
    \caption{PIs for Ridge with the UDBLag dataset with random seed 5678.}
    \label{fig:rg_permutation_inter_2}
    \end{subfigure}
    \caption{The FCs and PIs for Ridge classification
    with the 5G and the UDBLag dataset with two different random seeds.}
\label{fig:ridge_unified_feature_scores}
\end{figure*}
\begin{figure*}[ht]
    \centering
    \begin{subfigure}[t]{0.27\textwidth}
    \includegraphics[width=1\linewidth]{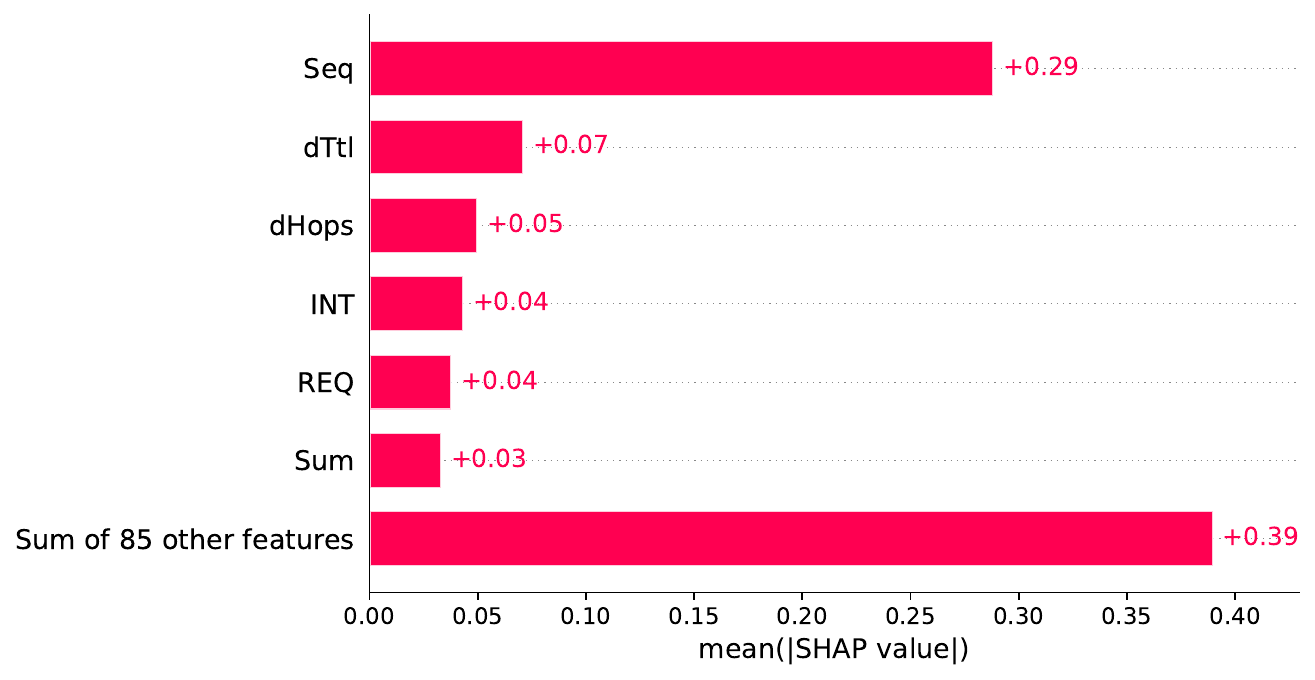}
    \caption{SHAPs for 5G\\dataset with seed 1234.}
    \label{fig:dnn_shap_5g_12345}
    \end{subfigure}\hfill
    \begin{subfigure}[t]{0.27\textwidth}
    \includegraphics[width=1\linewidth]{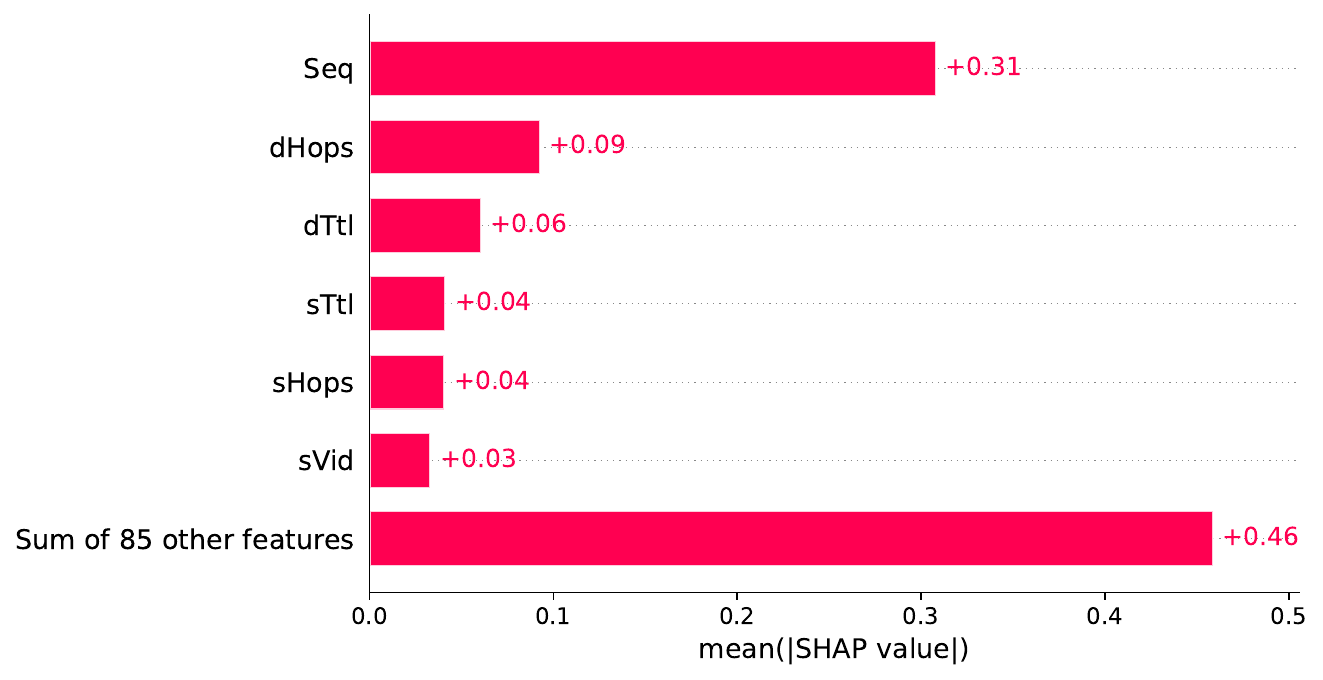}
    \caption{SHAPs for 5G\\dataset with seed 6789.}
    \label{fig:dnn_shap_5g_9876}
    \end{subfigure}\hfill
    \begin{subfigure}[t]{0.23\textwidth}
    \includegraphics[width=1\linewidth]{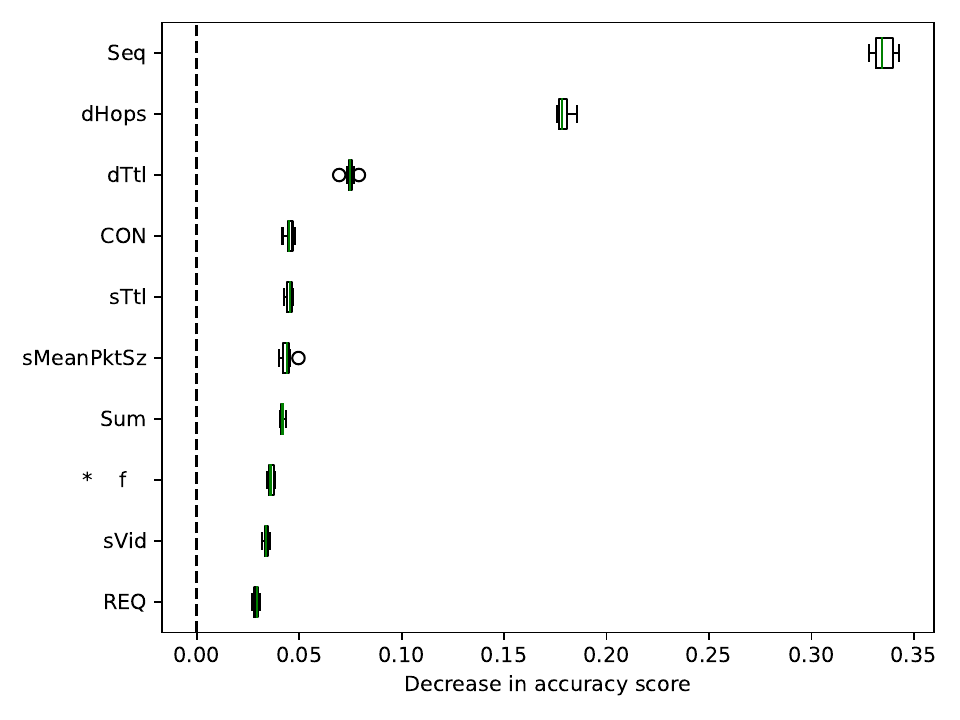}
    \caption{PIs for 5G dataset\\ with seed 1234.}
    \label{fig:dnn_per_5g_2_1234}
    \end{subfigure}\hfill
    \begin{subfigure}[t]{0.23\textwidth}
    \includegraphics[width=1\linewidth]{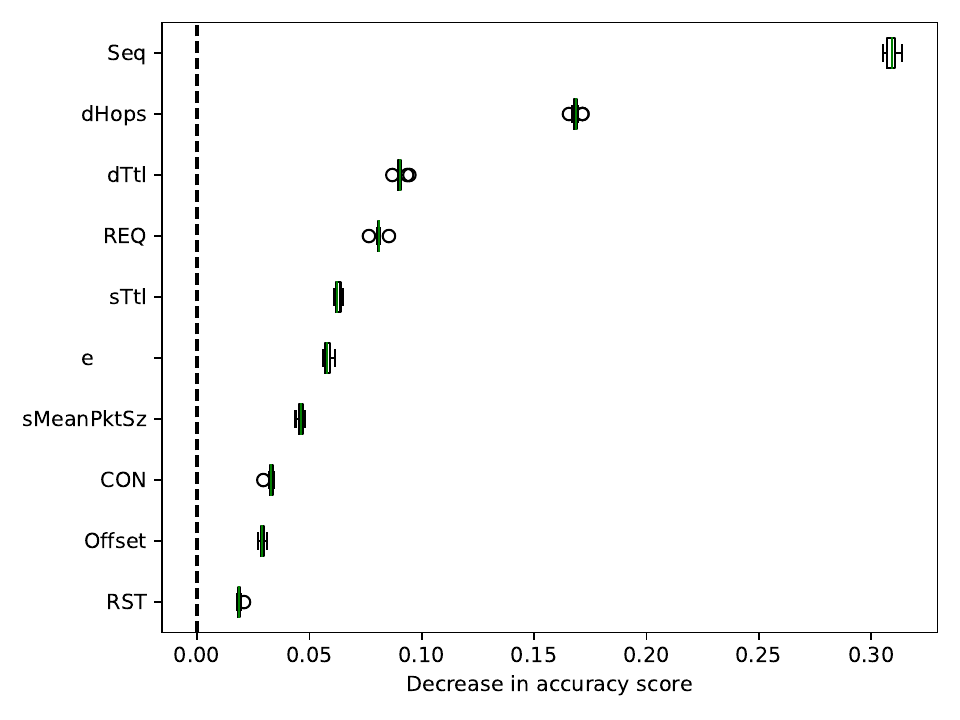}
    \caption{PIs for 5G dataset\\ with seed 6789.}
    \label{fig:dnn_per_5g5678}
    \end{subfigure}\hfill
    \begin{subfigure}[t]{0.26\textwidth}
    \includegraphics[width=1\linewidth]{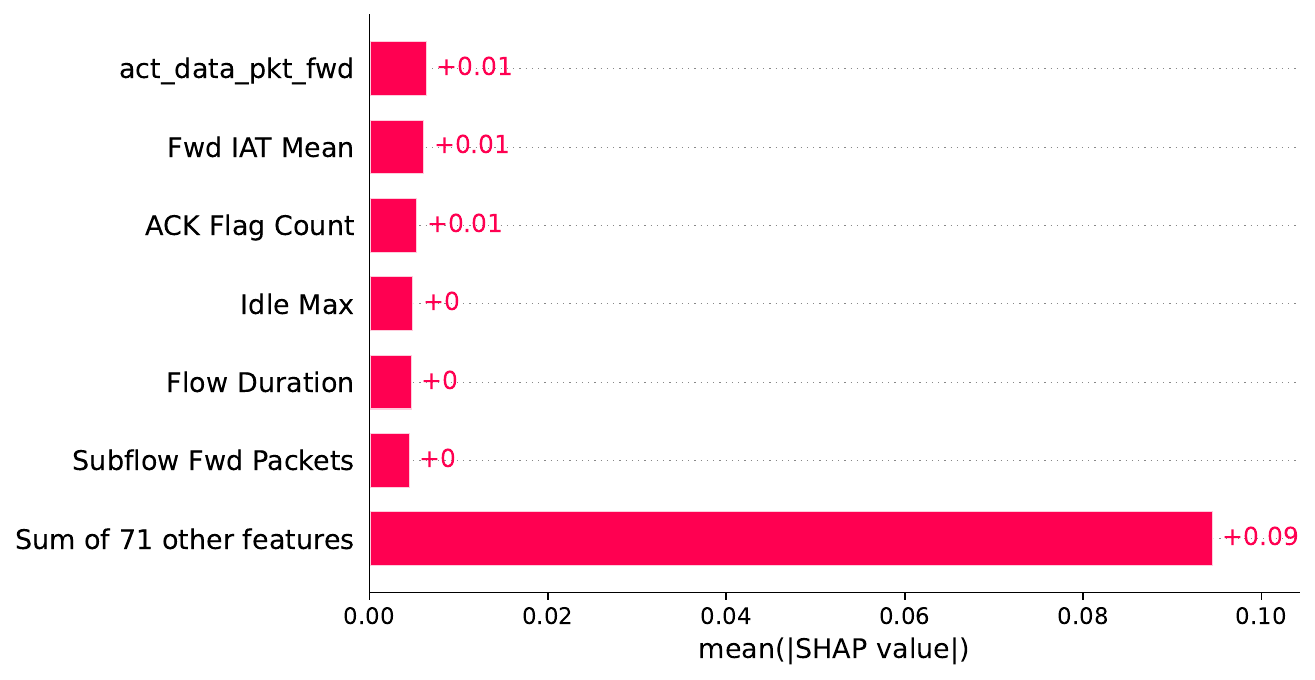}
    \caption{SHAPs for UDBLag\\with seed 1234.}
    \label{fig:dnn_shap_inter_12345}
    \end{subfigure}\hfill
    \begin{subfigure}[t]{0.26\textwidth}
    \includegraphics[width=1\linewidth]{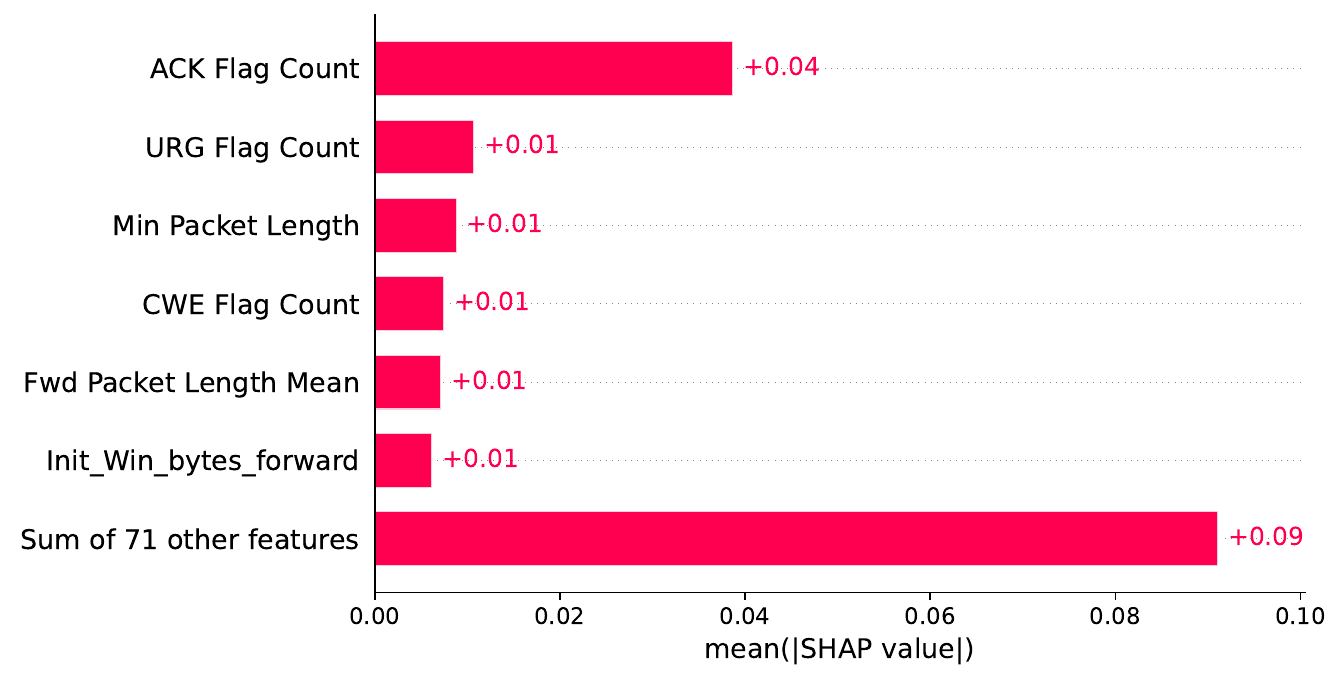}
    \caption{SHAPs for UDBLag\\ with seed 6789.}
    \label{fig:dnn_shap_inter_2_9876}
    \end{subfigure}\hfill
    \begin{subfigure}[t]{0.24\textwidth}
    \includegraphics[width=1\linewidth]{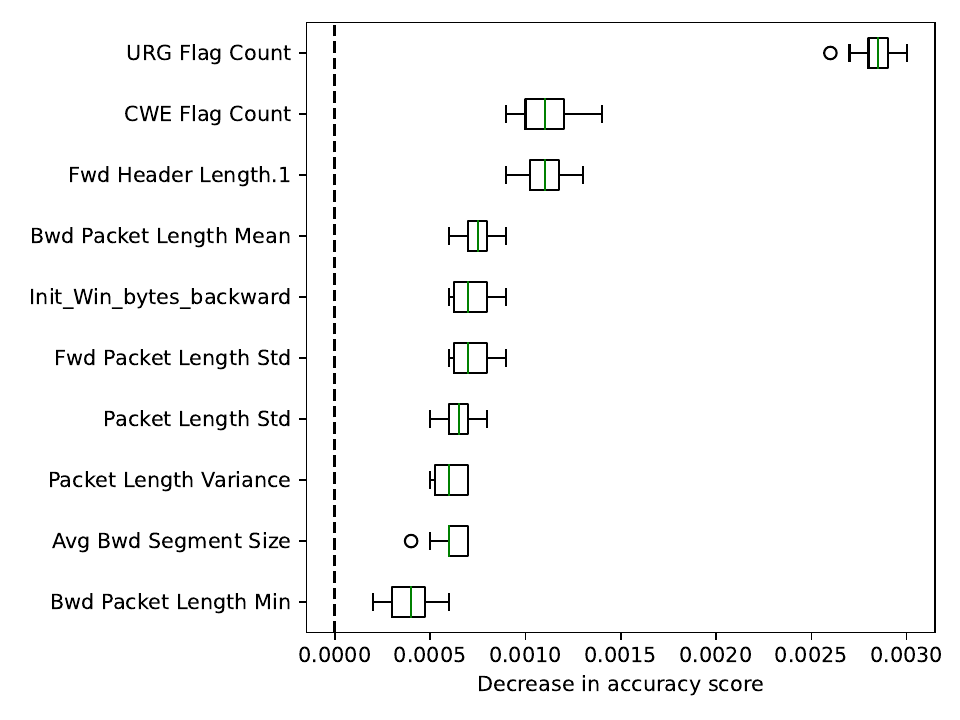}
    \caption{PIs for UDBLag\\ with seed 1234.}
    \label{fig:dnn_per_udblag_5678}
    \end{subfigure}\hfill
    \begin{subfigure}[t]{0.24\textwidth}
    \includegraphics[width=1\linewidth]{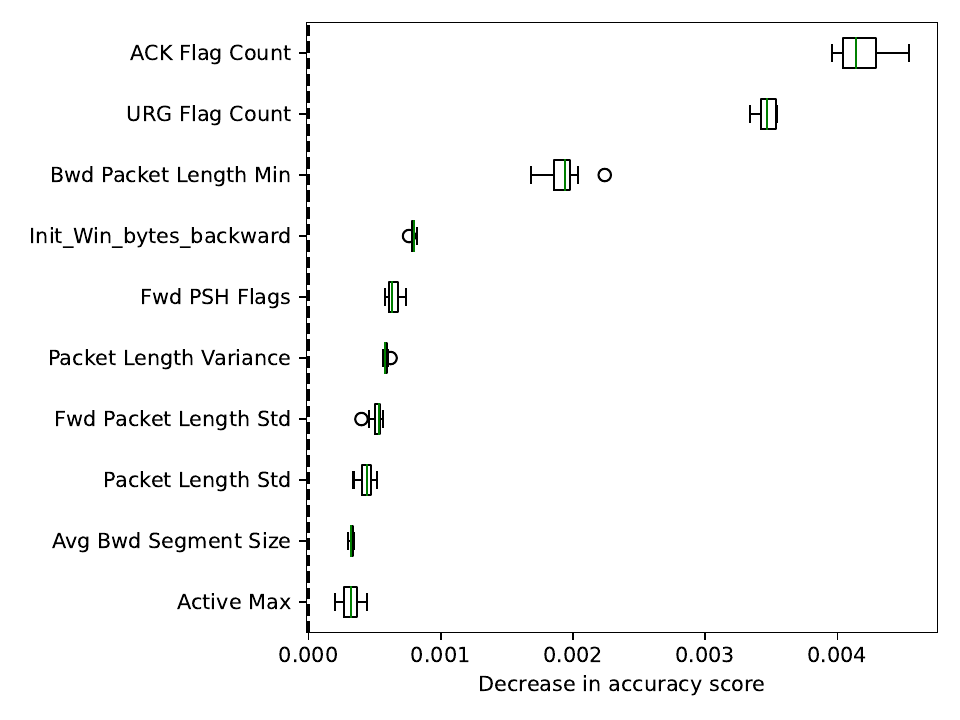}
    \caption{PIs for UDBLag\\ with seed 6789.}
    \label{fig:dnn_per_udblag_9876}
    \end{subfigure}
    \caption{The top features based on the SHAPs and PIs for DNN.}
    \label{fig:dnn_all_explanation_results}
\end{figure*}
Figure \ref{fig:ridge_unified_feature_scores} shows
the top features for Ridge classification for the 5G
and the UDBLag datasets. In particular, Figures \ref{fig:rg_feat_coefs_5g}
and \ref{fig:rg_feat_permutation_5g} show the FCs and the PIs for the 5G dataset.
We see that, as with DTs,
the top two features are \texttt{Seq} and \texttt{sTtl} and they do not seem to change 
with different random seeds. 
As for the UDBLag dataset, the results plotted in Figures \ref{fig:rg_feat_coefs_inter},
\ref{fig:rg_feat_coefs_inter2}, \ref{fig:rg_permutation_inter}, and 
\ref{fig:rg_permutation_inter_2} show that the top features
differ largely across different methods and random seeds.
As a note, since for linear models, SHAPs and FCs are equivalent, we only present FCs.

Figures \ref{fig:dnn_all_explanation_results} show
the top scoring features obtained by SHAPs and PIs for DNNs for the 5G
and the UDBLag datasets with two different seed values.
Figures \ref{fig:dnn_shap_5g_12345}, \ref{fig:dnn_shap_5g_9876}, \ref{fig:dnn_per_5g_2_1234} and
\ref{fig:dnn_per_5g5678} show that 
for the 5G dataset, \texttt{Seq} is consistently the most impactful feature affecting a DNN's prediction.
After \texttt{Seq}, \texttt{dTtL} and \texttt{dHops} are the two most impactful features for both SHAPs and PIs.
Moreover, considering all top features of Ridge classifiers, DTs, and DNNs, 
\texttt{Seq} is the only common top feature.
For the UDBLag dataset, Figures \ref{fig:dnn_shap_inter_12345}, \ref{fig:dnn_shap_inter_2_9876},
\ref{fig:dnn_per_udblag_5678} and \ref{fig:dnn_per_udblag_9876}
show that the most impactful features differ substantially across 
different training runs and external explanation methods. 

\colorlet{shadecolor}{red!10}
\begin{shaded*}
\textbf{Main Result \themycounter:}
\textit{Overall, our results indicate that there is
little to no consistency among the sets of the most impactful 
features across different ML models having intrinsic self-feedback, 
external explanation methods, and training runs
due to aleatoric (stochastic) uncertainty -- further discussed in Sections \ref{FeatureCorrelations} and \ref{LearningProcess}.
}

\end{shaded*} 
\stepcounter{mycounter}

\begin{figure*}[!ht]
    \centering
    \begin{subfigure}[t]{0.45\textwidth}
    \includegraphics[width=1\linewidth]{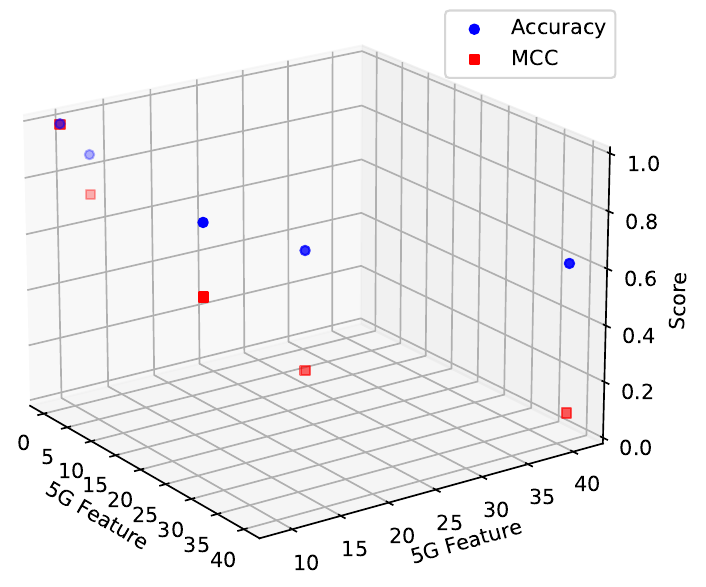}
    \caption{Average accuracy and MCC of DTs with the top cross-features with the 5G dataset.}
    \label{fig:dt_5g_cross_scores}
    \end{subfigure}\hfill
    \begin{subfigure}[t]{0.45\textwidth}
    \includegraphics[width=1\linewidth]{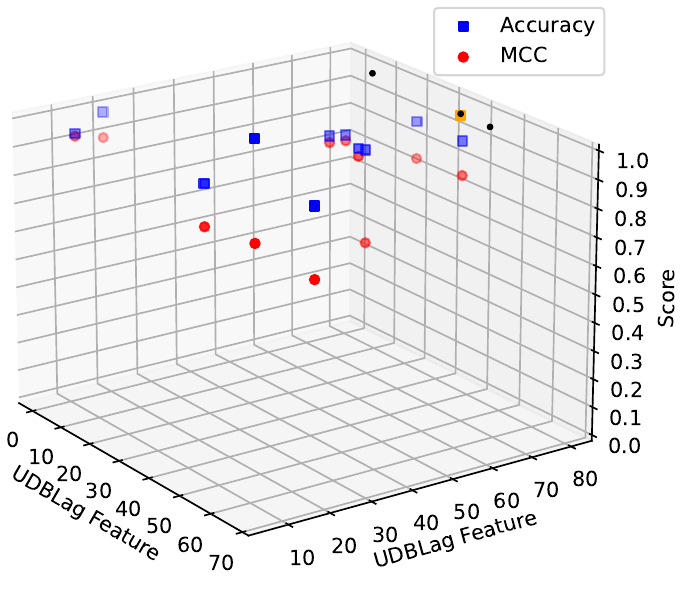}
    \caption{Average accuracy and MCC of DTs with the top cross-features with the UDBLag dataset.}
    \label{fig:dt_udblag_cross_scores}
    \end{subfigure}
    \caption{The pairs of the average accuracy and MCC of DTs with the top three cross-features. 
    To be able to visualize, we plot the top two features. For each pair, the accuracy is higher and plotted with a blue circle and the MCC is lower and plotted with a red square. Also, for each pair, 
    the accuracy and MCC are on the same line that is perpendicular to the feature plane.}
    \label{fig:dt_all_cross_scores}
\end{figure*}

\subsection{Cross-Explanations}
\label{CrossExplanations}
We now evaluate the impact of the top features that are obtained 
by different external explanation methods and classifiers
on the classification performance of a separate independent DT 
which is then trained by only using those top features.
We conduct this analysis to see 
whether the top features determined by different trained models and external methods
can be successful in other settings. That is, to probe if those top features are \textit{transferable}. 
This way of cross-evaluation of top features, which we name as \textit{cross-explanations},
provides a novel way of checking if the explanations, i.e. the most impactful features, are consistent
across different classifiers and settings. 

To decide whether a feature set is transferable, we use a threshold of $MCC >= 0.95$. We note that
if $MCC >= 0.95$ for a 2x2 confusion matrix, then all other binary classification metrics are guaranteed 
to be $>= 0.95$. We say a set of features is \textit{transferable} if an ML model that is only trained with
the feature set achieves an $MCC >= 0.95$.
In our analysis, we limit the set of top features to have three features so that we are 
 able to determine the impact of a feature set in a nuanced way.

Figure \ref{fig:dt_5g_cross_scores} shows the pairs of the average accuracy and MCC of DTs 
that are trained with the top three features obtained by different external methods 
or classifiers for the 5G dataset. 
Specifically, the five sets of top three features in Figure \ref{fig:dt_5g_cross_scores} are obtained 
 \begin{itemize}
\item by the highest three FCs of a Ridge classifier,  
\item by the highest three SHAPs of a DNN,
\item by the highest three SHAPs of a separate and independently trained DT,
 \item the highest three PIs of a separate and independently trained DT, and
 \item the highest three FIs of a separate and independently trained DT.
 \end{itemize}
 We choose to plot only the standard accuracy metric and MCC for a clear presentation. 
From Figure \ref{fig:dt_5g_cross_scores}, we see that among the five sets, 
only one set is transferable (the leftmost point) for the 5G dataset. 
Similarly, for the UDBLag dataset, we train and evaluate DTs 
with the top three features determined by different external methods and classifiers. 
As shown in Figure \ref{fig:dt_udblag_cross_scores}, 
we report eleven sets of top three features. 
These eleven sets of top three features consist of:
\begin{itemize}
     \item two sets having the highest three FCs of two different Ridge classifiers,
     \item two sets having the highest three SHAPs of two different DNNs,
     \item two sets having the highest three PIs of two different DNNs,
     \item two sets having the highest three SHAPs of two separate and independently trained DTs, 
     \item two sets having the highest three PIs of two separate and independently trained DTs, and
    \item one set having the highest three FIs of two separate and independently trained DTs.
\end{itemize}
Figure \ref{fig:dt_udblag_cross_scores} shows that among the eleven sets of top three features, only four are transferable. In Figure \ref{fig:dt_udblag_cross_scores}, to improve the visualization,  
we also plot the projections of the MCCs of those four sets 
onto the right plane, where three projections are black dots 
and one is an orange square. We use an orange square to make 
the two very close MCCs visibly distinguishable. 
\begin{shaded*}
\textbf{Main Result \themycounter:}
\stepcounter{mycounter}
\textit{To summarize, given a fixed input dataset, 
the classification performance of a set of top features that are determined to 
be influential by external explanation methods or intrinsic self-feedback of ML models 
is not necessarily achieved in other settings where the model and/or the feature set (for training) is different. This, in turn, corroborates the absence of consistent explanations across different settings.}
\end{shaded*}

\begin{figure*}[!ht]
    \centering
    \begin{subfigure}[t]{0.31\textwidth}
    \includegraphics[width=1\linewidth]{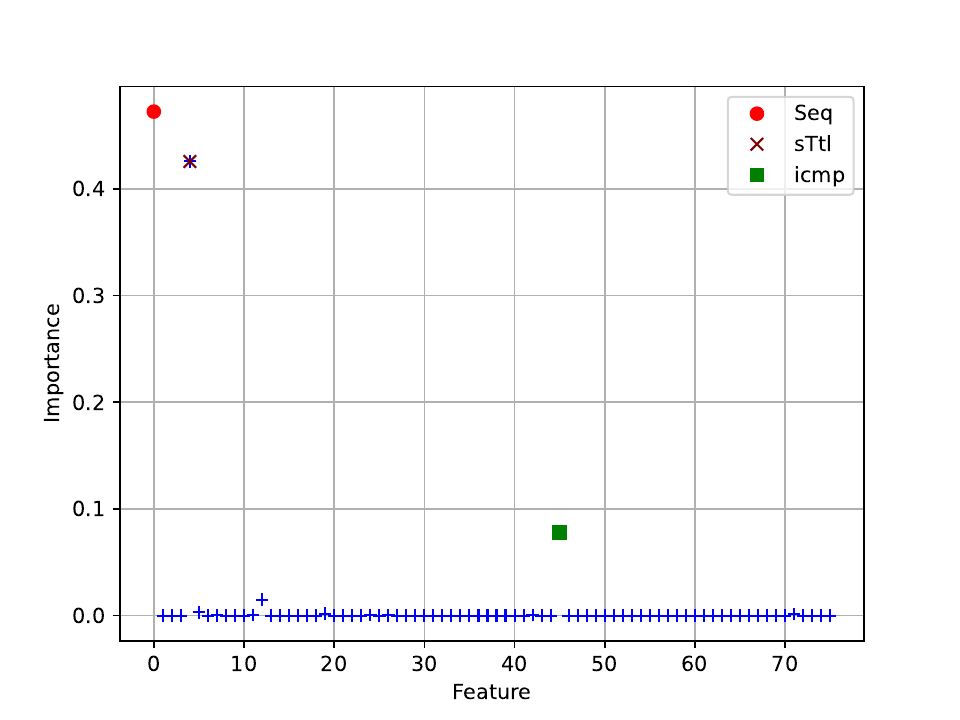}
    \caption{FIs for the 5G Dataset.}
    \label{fig:correlations_dt_importances_5g_1234}
    \end{subfigure}\hfill
    \centering
    \begin{subfigure}[t]{0.34\textwidth}
    \includegraphics[width=1\linewidth]{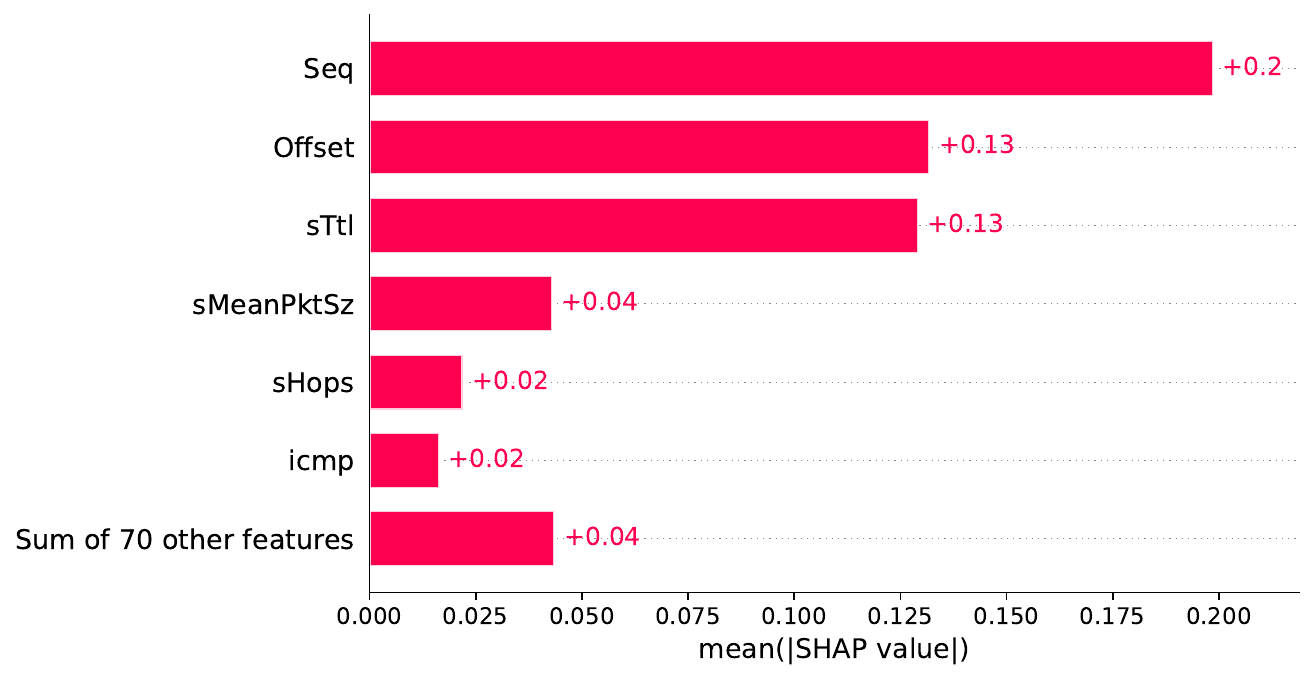}
    \caption{SHAPs for the 5G Dataset.}
    \label{fig:correlations_dt_shap_5g_1234}
    \end{subfigure}\hfill
    \centering
    \begin{subfigure}[t]{0.31\textwidth}
    \includegraphics[width=1\linewidth]{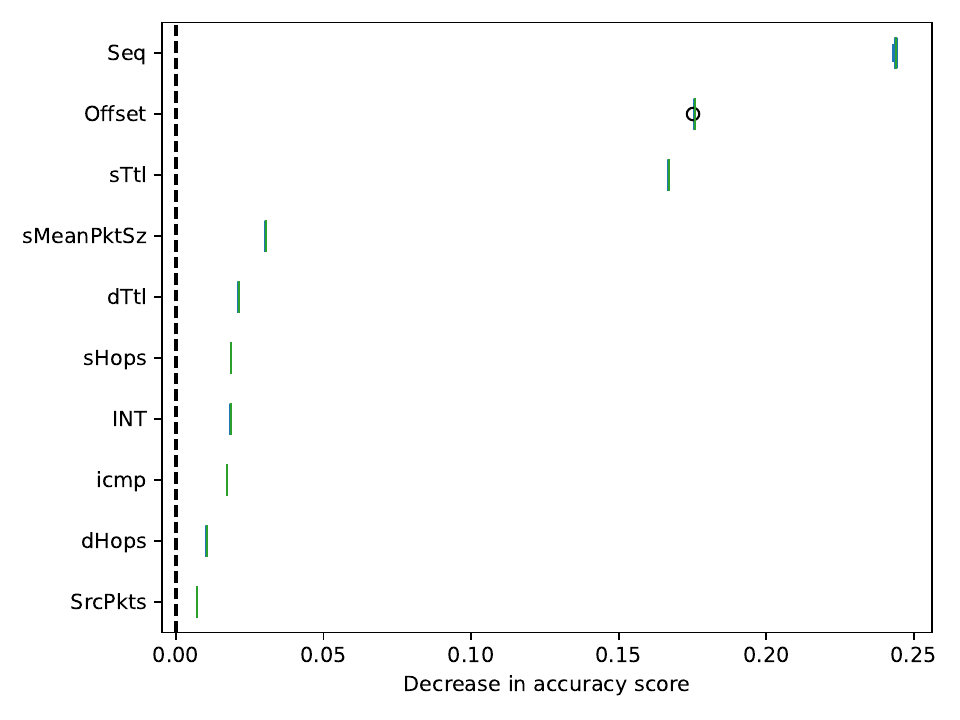}
    \caption{PIs for the 5G Dataset.}
    \label{fig:correlations_dt_permutations_5g_1234}
    \end{subfigure}\hfill
    \centering
    \begin{subfigure}[t]{0.31\textwidth}
    \includegraphics[width=1\linewidth]{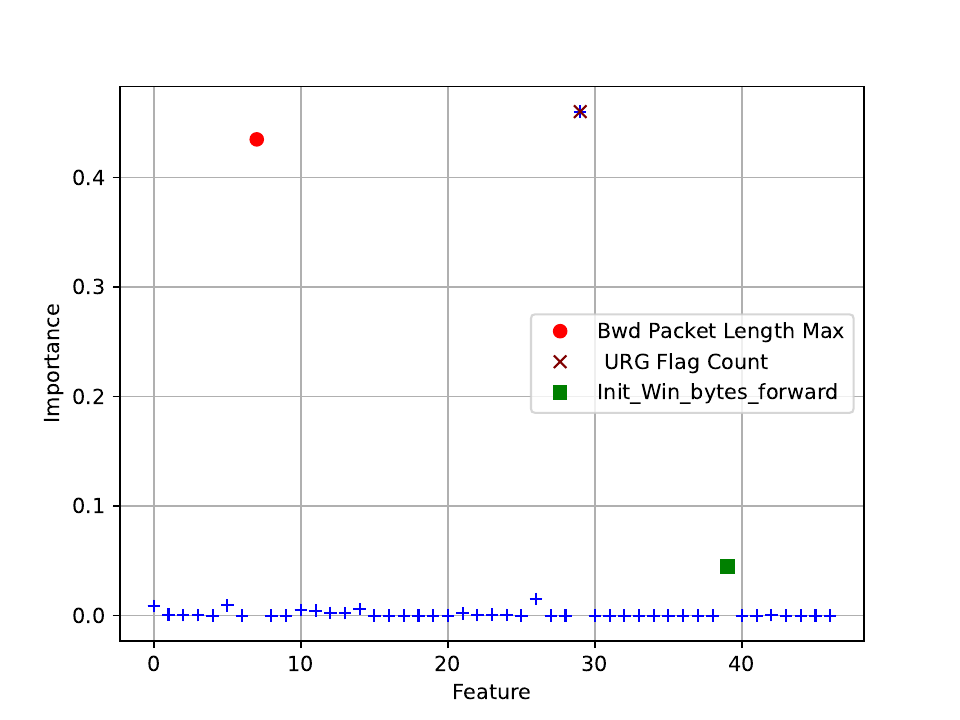}
    \caption{FIs for the UDBLag\\ Dataset.}
    \label{fig:correlations_dt_importances_udblag1234}
    \end{subfigure}\hfill
    \centering
    \begin{subfigure}[t]{0.34\textwidth}
    \includegraphics[width=1\linewidth]{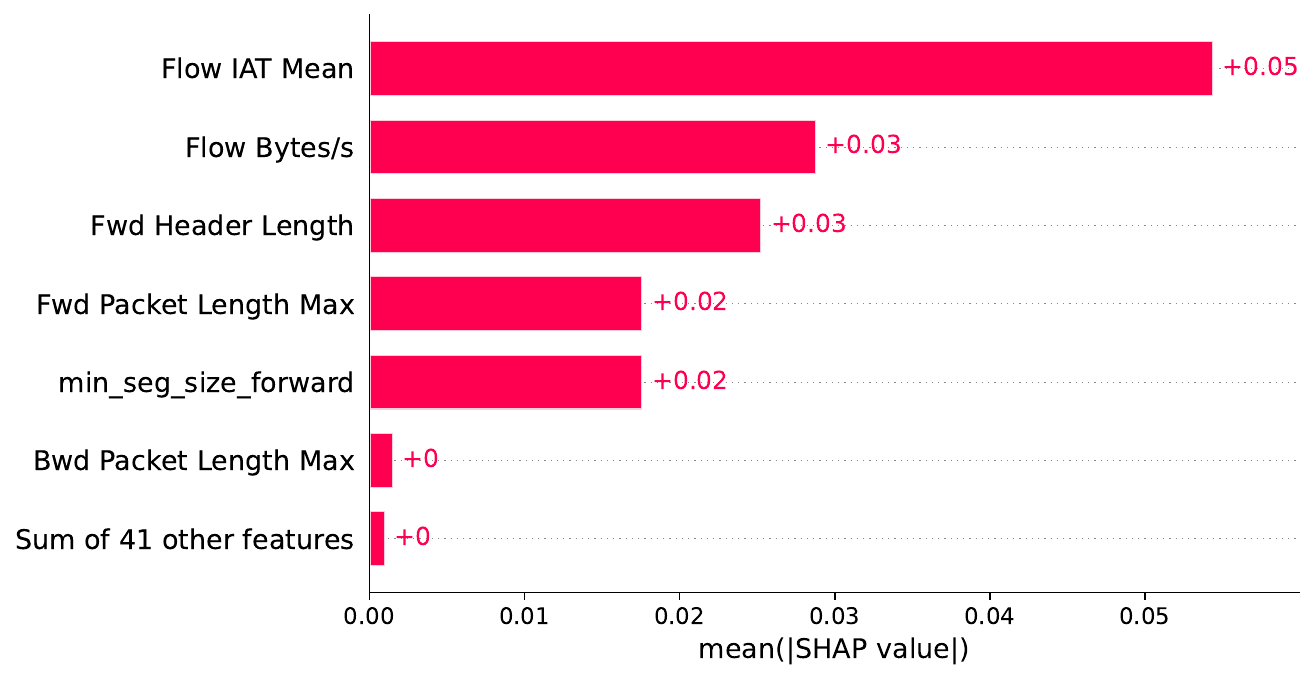} 
    \caption{SHAPs for the UDBLag\\ Dataset.}
    \label{fig:correlations_dt_shap_udblag_1234}
    \end{subfigure}\hfill
    \centering
    \begin{subfigure}[t]{0.31\textwidth}
    \includegraphics[width=1\linewidth]{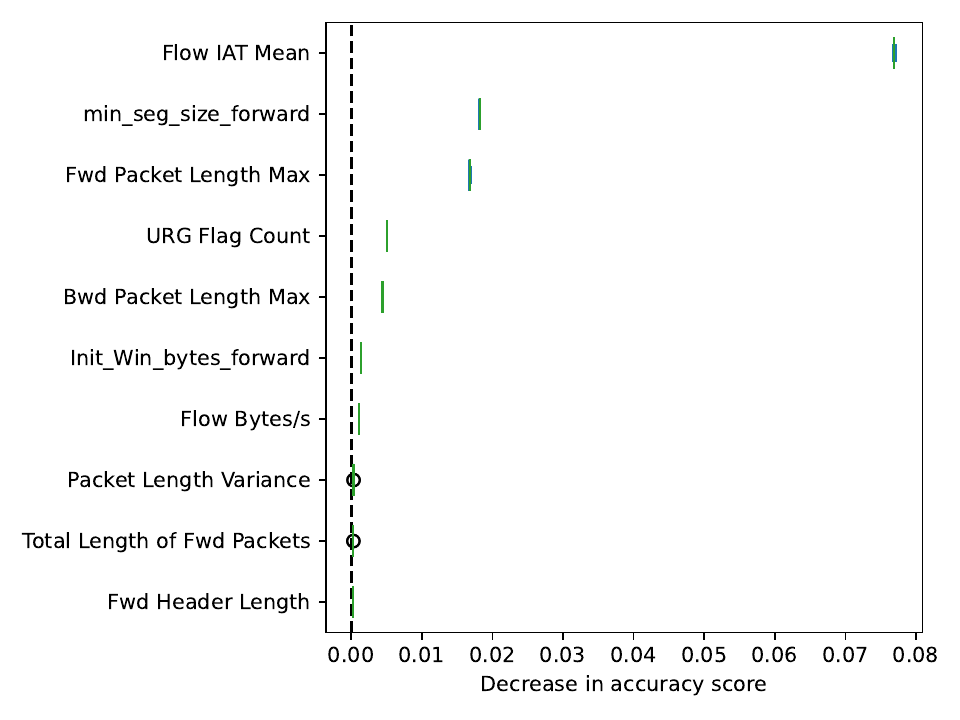} 
    \caption{PIs for the UDBLag\\Dataset.}
    \label{fig:correlations_dt_permutations_udblag_1234}
    \end{subfigure}
\caption{The top features based on the FIs, SHAPs, and PIs for 
DTs after the correlated features omitted.}
\label{fig:dt_correlation_results}
\end{figure*}
\begin{figure*}[ht]
    \centering
    \begin{subfigure}[t]{0.27\textwidth}
    \includegraphics[width=1\linewidth]{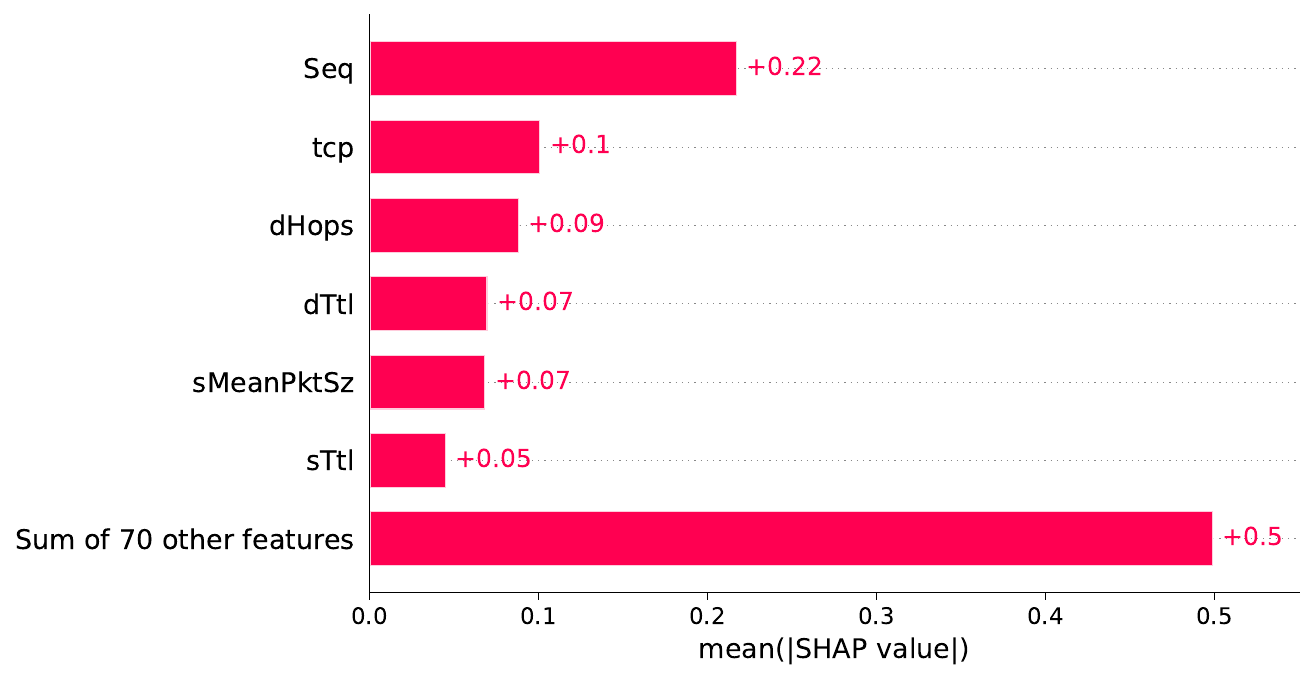}
    \caption{SHAPs for the 5G\\ Dataset.}
    \label{fig:correlations_dnn_shap_5g_12345}
    \end{subfigure}\hfill
    \begin{subfigure}[t]{0.22\textwidth}
    \includegraphics[width=1\linewidth]{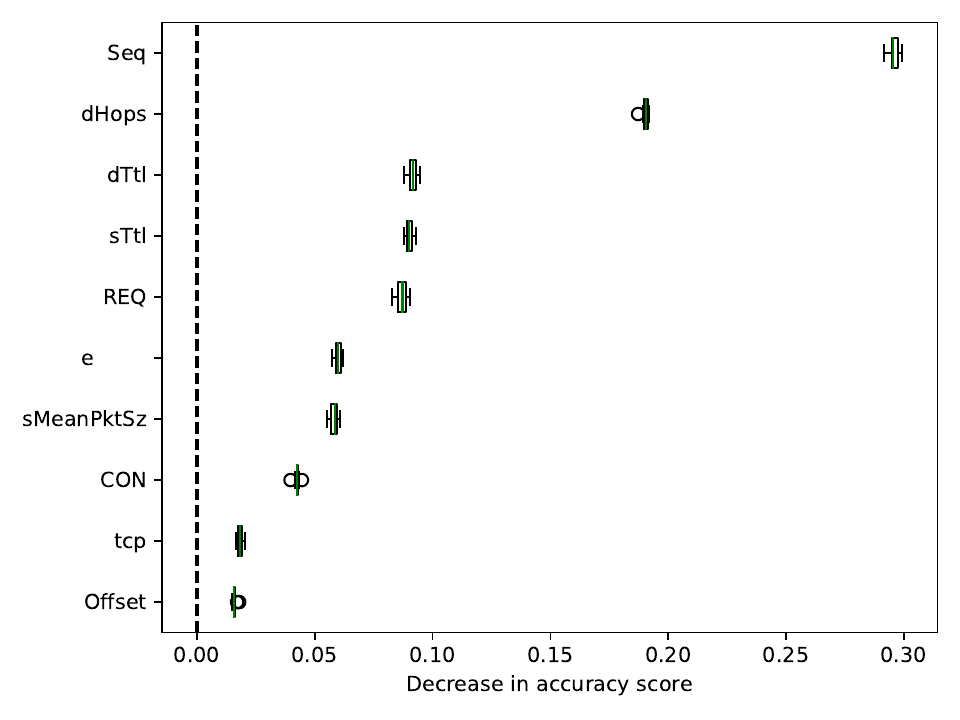}
    \caption{PIs for the 5G\\ Dataset.}
    \label{fig:correlations_dnn_permutations_5g_12345}
    \end{subfigure}\hfill
    \begin{subfigure}[t]{0.27\textwidth}
    \includegraphics[width=1\linewidth]{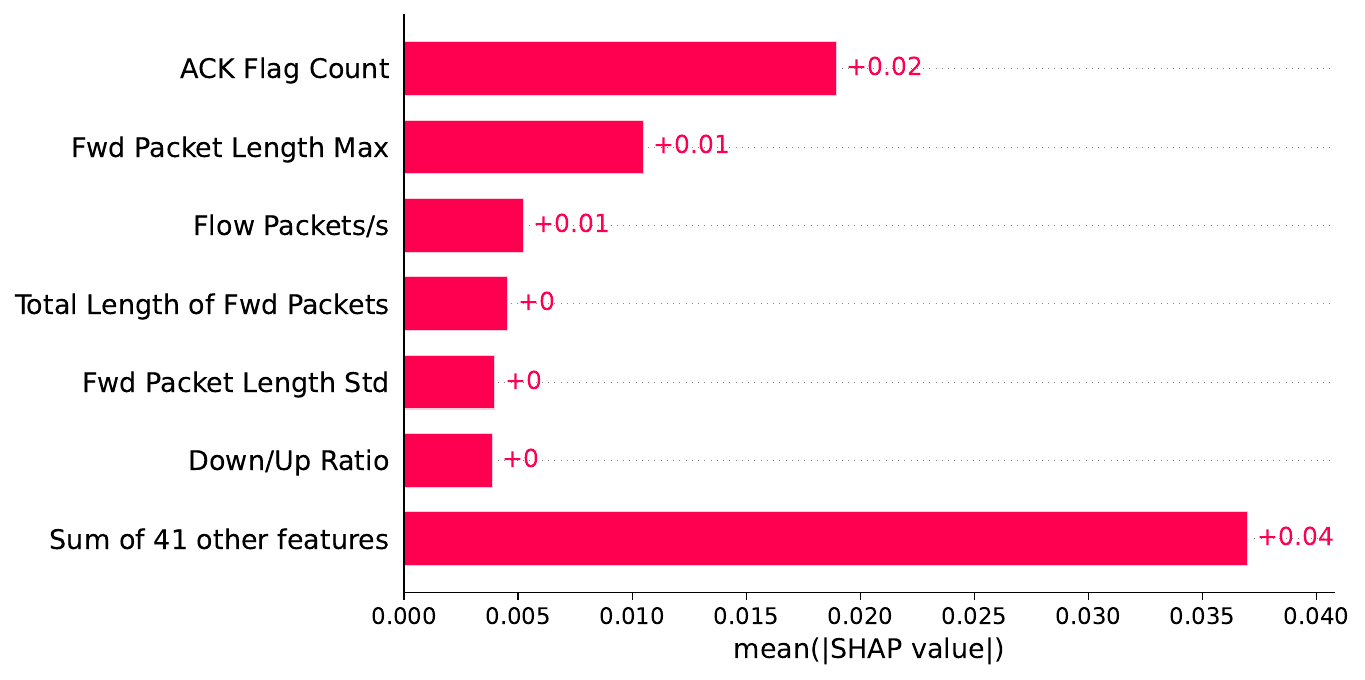}
    \caption{SHAPs for the\\ UDBLag Dataset.}
    \label{fig:correlations_dnn_shap_udblag_1234}
    \end{subfigure}\hfill
    \begin{subfigure}[t]{0.23\textwidth}
    \includegraphics[width=1\linewidth]{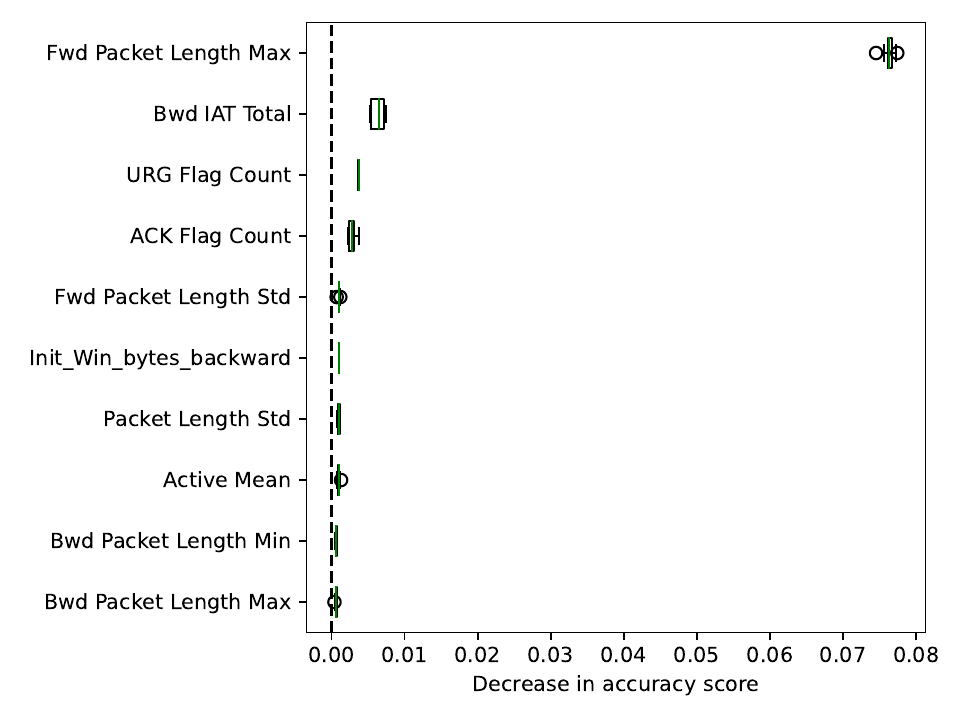}
    \caption{PIs for the\\ UDBLag Dataset.}
    \label{fig:correlations_dnn_permutations_udblag_1234}
    \end{subfigure}
    \caption{The top features based on the SHAPs and PIs for DNNs after the correlated features omitted.}
    \label{fig:dnn_correlation_explanation_results}
\end{figure*}
\subsection{Feature Correlations}
\label{FeatureCorrelations}
As we stated earlier, some external explanation methods, such as PIs and SHAPs, 
assume the independence of data features. 
In the cases where the features are not independent, that is, correlated,
such methods might mislead. There might be cases where
correlation causes the computed importance/impact 
of the correlated features to be lower and other features to be higher.
This, in effect, makes all computed importances unreliable.
For instance, for PIs,
if two features are strongly correlated and one of the features is permuted, 
an ML model still has access to the permuted feature through its correlated feature. 
This may cause lower PIs for both features and 
at the same time may falsely inflate PIs for other features. 
In other cases, an external method may falsely show that all features are unimportant. 
However, there are actually important/impactful features (see \cite{sklearn-corr} for a specific example). 

In our correlation study, we consider a feature correlation that is $>= 0.95$ as a strong correlation.
To find the features that have at least one strong correlation with another feature, 
we compute the correlation matrix based on Pearson's
correlation. Then, if a feature has a strong correlation with at least one other feature, 
we omit the feature. For the 5G dataset,
we find that there are 15 features that are strongly correlated with at least one other feature
out of 91 features. 
For the UDBLag dataset, there are 30 such features out of 77 features. 
As the next step, we omit strongly correlated features and then compute SHAPs and PIs for 
DTs and DNNs which are trained with only the remaining features. 

We note that when some features are omitted, 
the intrinsic explanations, i.e., FIs of DTs 
and FCs of Ridge, change by definition. 
In addition, as noted above, for linear models, 
SHAPs and FCs are equivalent, which means SHAPs 
for Ridge will change by definition as well. 
As a result, we do not discuss Ridge separately.
 
Figure \ref{fig:dt_correlation_results} shows the results for DTs.
Compared to the results with all features in Figure \ref{fig:5g_unified_feature_scores}
for the 5G dataset, we see that since the feature \texttt{ECO} is omitted, 
the set of top impactful features does not include it as in \ref{fig:correlations_dt_importances_5g_1234},
\ref{fig:correlations_dt_shap_5g_1234} and \ref{fig:correlations_dt_permutations_5g_1234}.
For the UDBLag dataset, because \texttt{Min Packet Length} and \texttt{Max Packet Length} are omitted, 
they are no longer among the top impactful features for DTs as in \ref{fig:correlations_dt_importances_udblag1234},
\ref{fig:correlations_dt_shap_udblag_1234} and \ref{fig:correlations_dt_permutations_udblag_1234}
as opposed to the results with all features in Figure \ref{fig:dt_udblag_unified_feature_scores}. 
Additionally, since \texttt{Fwd IAT Total} and \texttt{Fwd IAT Mean} are omitted, they are no longer
among the top impactful features as opposed to the top features in Figure \ref{fig:dt_shapley_udb_5678}.
Figure \ref{fig:dnn_correlation_explanation_results} shows the correlation results for DNNs. 
We see that there are no top features that are 
also strongly correlated for the 5G dataset.
For the UDBLag dataset, since the features \texttt{Fwd IAT Mean} and \texttt{Min Packet Length}
are omitted, they are no longer among the top impactful features in contrast with the results
with all features in Figures \ref{fig:dnn_shap_inter_12345} and \ref{fig:dnn_shap_inter_2_9876}.
\colorlet{shadecolor}{red!10}
\begin{shaded*}
\textbf{Main Result \themycounter:}
\stepcounter{mycounter}
\textit{When we omit strongly correlated features to prevent 
computing potentially inaccurate and unreliable feature importances, 
we may end up with new sets of the top impactful features that are inconsistent 
with the top features when all features are considered. 
Consequently, feature correlation might become yet another source of uncertainty exacerbating the problem of 
inconsistent explanations.}
\end{shaded*}

\subsection{Constituents of a Learning Process}
\label{LearningProcess}
A learning process for an ML model refers to all aspects of
data pre-processing steps, such as data cleaning, standardization, feature selection \cite{subasi2024analysis}, 
and dimensionality
reduction, as well as the training phase where model and hyper-parameter optimizations are performed. 
We have already explored the interplay between some of these aspects and the computed explanations above. 
For instance, as a feature selection step, we have identified strongly correlated features and omitted them
from the training data. In this section, we further investigate the impact of other aspects of the  
learning process, namely, hyper-parameters and the optimization routine utilized during training.
While we do not report the results for all aspects of a learning process for brevity, 
our conclusions applies to them too. 

\subsubsection{Hyper-parameters:}
\begin{figure*}[!ht]
    \centering
    \begin{subfigure}[t]{0.5\textwidth}
    \includegraphics[width=1\linewidth]{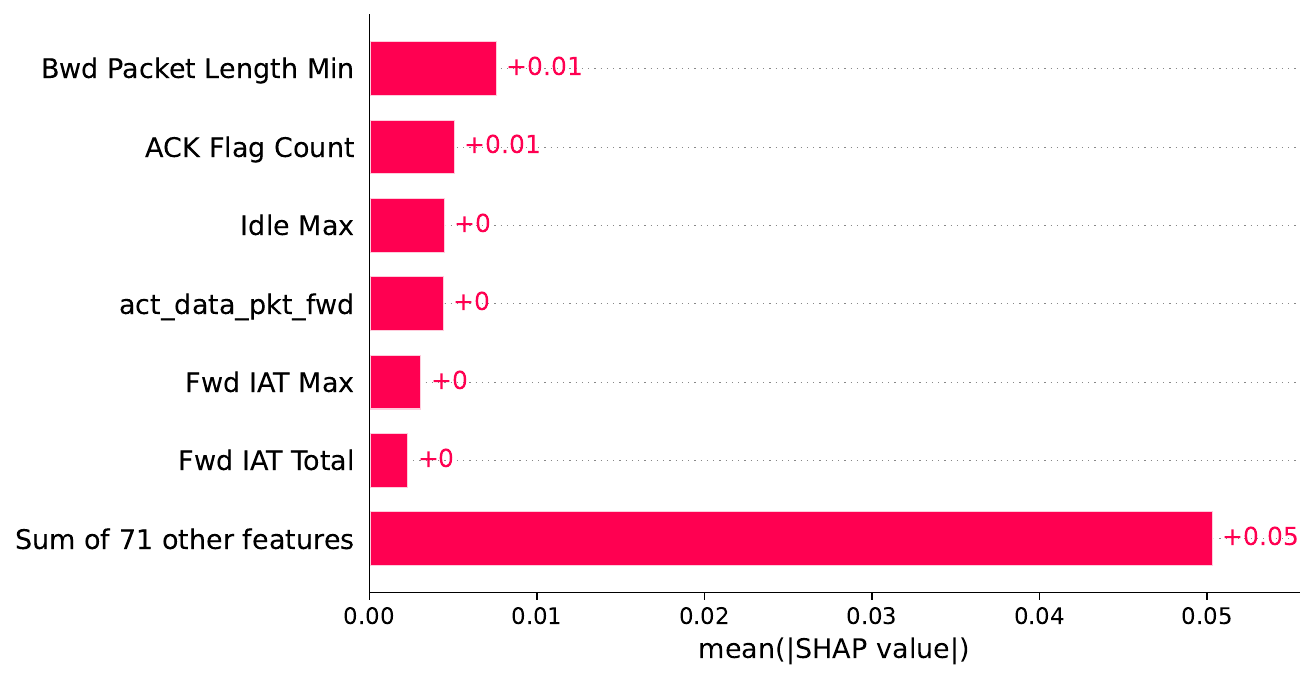}
    \caption{SHAPs with a batch size of 512.}
    \label{fig:dnn_shap_inter_batch_size_6789}
    \end{subfigure}\hfill
    \begin{subfigure}[t]{0.5\textwidth}
    \includegraphics[width=1\linewidth]{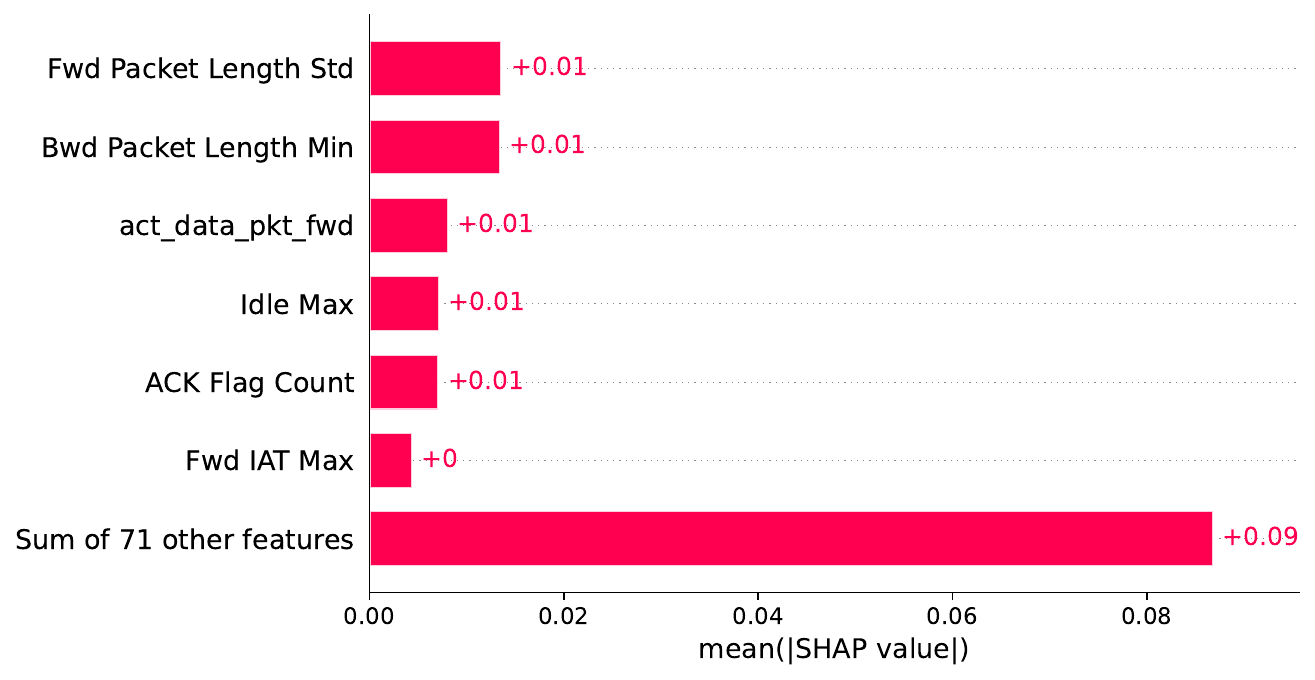}
    \caption{SHAPs with a 75\%-25\% train-test split.}
    \label{fig:dnn_shap_inter_split_ratio_25_6789}
    \end{subfigure}\hfill
    \caption{Top features based on SHAPs of a DNN for the UDBLag Dataset.}
    \label{fig:dnn_hyperparameters}
\end{figure*}
\begin{figure*}[!ht]
    \centering
    \begin{subfigure}[t]{0.47\textwidth}
    \includegraphics[width=1\linewidth]{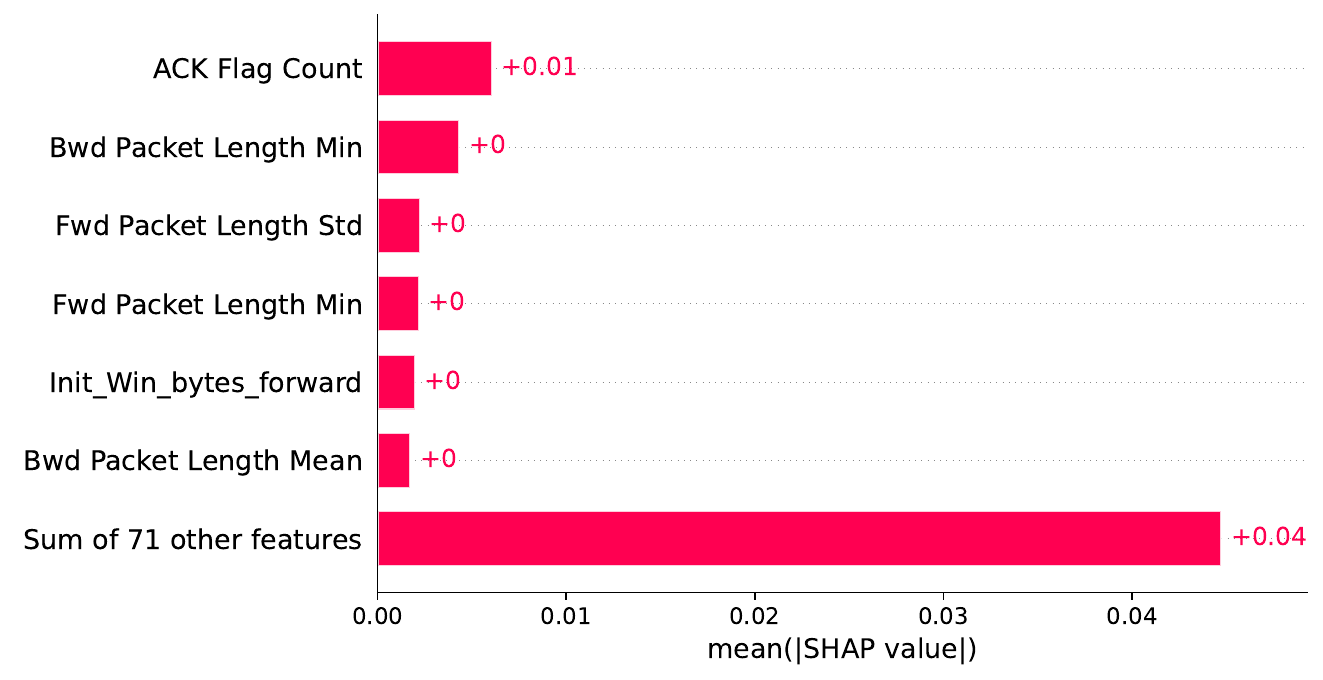}
    \caption{SHAPs with \texttt{Adam} optimizer instead of the default \texttt{RMSprop}.}
    \label{fig:dnn_shap_inter_optimizer_adam_25_1234}
    \end{subfigure}\hfill
    \begin{subfigure}[t]{0.47\textwidth}
    \includegraphics[width=1\linewidth]{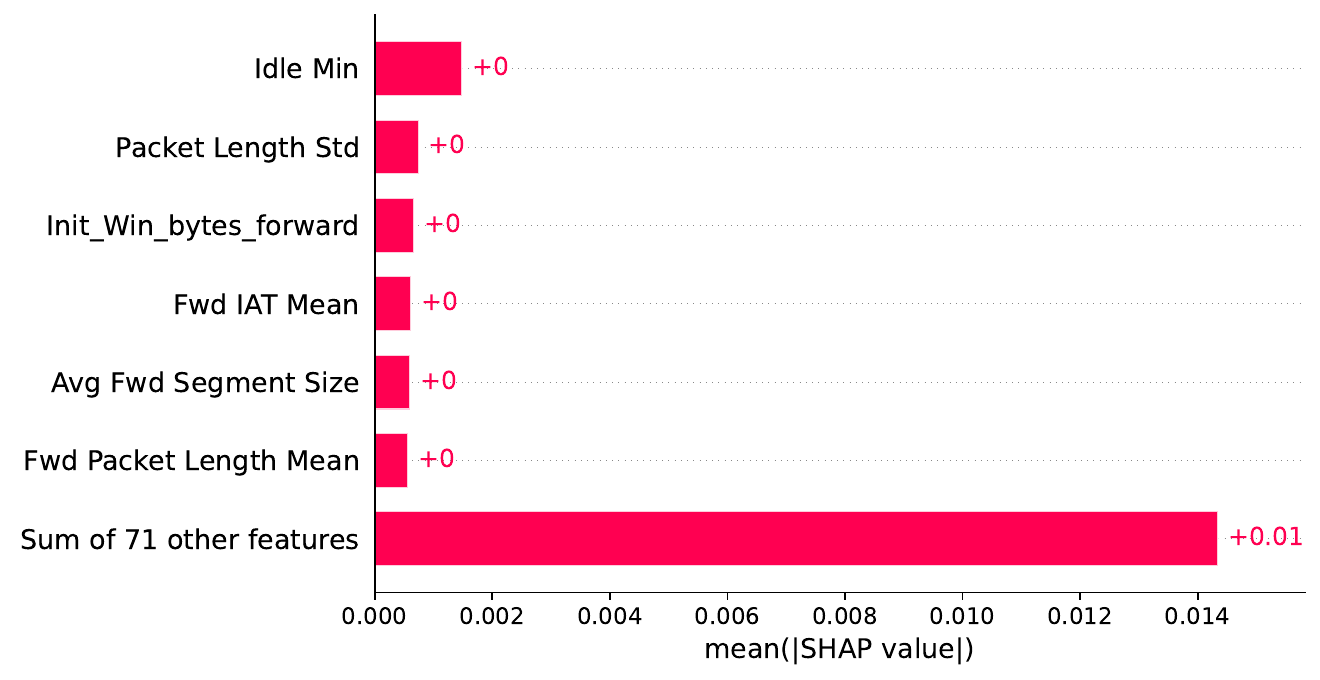}
    \caption{SHAPs with \texttt{Adam} optimizer, batch size of 512 and a 75\%-25\% train-test split.}
    \label{fig:dnn_shap_inter_optimizer_adam_15_512_1234}
    \end{subfigure}\hfill
    \caption{Top features based on SHAPs of a DNN for the UDBLag Dataset.}
    \label{fig:dnn_optimization_methods}
\end{figure*}

Most ML models and algorithms have hyper-parameters that specify and control the learning process
as opposed to the parameters of an ML model which are optimized during training.
Examples include train-test split ratio, batch size, learning rate, the number of iterations, and 
the number of (hidden) layers in a DNN. 
In this section, we run experiments to check 
if an explanation for a model changes when hyper-parameter values change.
Ideally, value changes should not influence model explanations. However, our results show that
they indeed do affect the resulting explanations.
In our evaluation, we change the hyper-parameters of batch size and train-test split ratio 
for a DNN while keeping other hyper-parameters the same as they are specified in Subsection \ref{classificationperfs}. 
We choose SHAPs as the explanation method for a DNN for the UDBLag Dataset.
Compared to the SHAPs in Figures \ref{fig:dnn_shap_inter_12345}, \ref{fig:dnn_shap_inter_2_9876}, 
and \ref{fig:correlations_dnn_shap_udblag_1234},
Figure \ref{fig:dnn_shap_inter_batch_size_6789} shows that the top features are different 
when we use a batch size of 512 instead of 256 which is used in all previous evaluations of DNNs above.
This is also true for the train-test split ratio. Figure \ref{fig:dnn_shap_inter_split_ratio_25_6789}
shows the SHAPs when 75\%-25\% train-test split ratio is used instead of 85\%-15\% in the earlier evaluations.
While we only present batch size and train-test split ratio to demonstrate that when their values change, SHAP
explanations change, this holds true for other hyper-parameters, classifiers, and explanation methods.  

\subsubsection{Optimization Methods:}
ML algorithms and methods generally employ different optimization
techniques. Moreover, these algorithms typically are amenable to 
different techniques/routines. 
While different optimization routines may have different convergence 
speed and characteristics, they preferably should
not affect the final model and its performance. 
To check this, we perform experiments to observe the effect of employing
different optimizers on model explanations.
Figure \ref{fig:dnn_shap_inter_optimizer_adam_25_1234}
shows the resulting SHAPs when the default Keras optimizer \texttt{RMSprop} 
is replaced with the \texttt{Adam} optimizer to train a DNN for the UDBLag Dataset. 
When compared with Figures \ref{fig:dnn_shap_inter_12345}, \ref{fig:dnn_shap_inter_2_9876}, 
and \ref{fig:correlations_dnn_shap_udblag_1234}, and \ref{fig:dnn_hyperparameters},
we see that the top impactful features do not match, that is, the explanations are different 
and inconsistent. As a further evaluation step, 
we conduct experiments with the \texttt{Adam} optimizer, a batch size of 512, and 
a 75\%-25\% train-test split to see how the changes in 
hyper-parameters and in the optimization routine 
compound. Figure \ref{fig:dnn_shap_inter_optimizer_adam_15_512_1234}
shows the results. Compared with
Figure \ref{fig:dnn_shap_inter_optimizer_adam_25_1234}, Figure
\ref{fig:dnn_shap_inter_optimizer_adam_15_512_1234}
shows that the \texttt{Init\_Win\_bytes\_forward} is the only shared top feature between the two explanations. 
This suggests that combining multiple changes could severely degrade 
that the stability, the consistency, and the reliability of explanations. 
Ideally, such changes in hyper-parameters and the optimization technique 
should not drastically impact model explanations.

\begin{shaded*}
\textbf{Main Result \themycounter:}
\stepcounter{mycounter}
\textit{In all experiments reported in this subsection, 
the classification performances do not
differ from each other for more than 0.1\% for the standard classification metrics and 2\% for MCC when different
hyper-parameter values or optimization routines are tested. These limited performance variations are critical:
We see that while the classification performances essentially remain the same, the explanations do not.
As a result, this observation discredits the merit of feature-based model explanations.}  
\end{shaded*}

\subsection{Lack of Actionability and Utility}
\label{Actionability}
One of the ultimate goals of interpretable and explainable ML is to
be able to identify and provide necessary actions if there are issues regarding the
model under consideration. However, it is not clear that what actions are needed to be taken
if the explanation is a list of the most important/impactful features based on some definition of importance.
How does it help to know that the top most important features are, say, 
\texttt{Idle Min, Packet Length Std, Init\_Win\_bytes\_forward, Fwd IAT Mean and Avg Fwd Segment Size}?
A list of most important features can not help much in deciding what to do next after a cyber attack.
The situation worsens when a DNN or any other black-box model
is employed to detect intrusions. This is because they do not offer any feedback other than
the feature importances provided by an external explanation method.
In comparison, DTs intrinsically provide two types of feedback for their predictions. 
These are FIs and the boolean logic rules obtained by traversing from the root to the output 
leaf node and conjoining the conditions along the path. 
The logic rules also serve as a mechanism that reveals how the features
interact with each other. However, having said that, 
the rampant stochastic inconsistencies in these feature-based explanations nullify the value of DTs' intrinsic feedback. 
In consequence, current interpretable and explainable ML methods 
are not able to recommend sound and effective after-the-fact actions in cybersecurity 
and more specifically in intrusion detection. 
\colorlet{shadecolor}{red!10}
\begin{shaded*}
\textbf{Main Result \themycounter:}
\stepcounter{mycounter}
\textit{
Considering the stochastic inconsistencies arising from random variations and the changes 
in the constituents of a learning process, and their inability to provide actionable feedback, the state of the art
feature-based explanations - obtained intrinsically or externally - exhibit no tangible practical value 
for cybersecurity experts.}
\end{shaded*}

\colorlet{shadecolor}{green!10}
\begin{shaded*}
\textbf{Guide \themyadvise:}
\stepcounter{myadvise}
\textit{
In conclusion, we advise that research studies on interpretable and explainable ML 
for intrusion detection should focus on the
approaches that are not based on feature impact or importance. 
Potential approaches include prototype-based models and counterfactual explanations. 
}
\end{shaded*}

\section{Conclusion and Future Work}
\label{conclusion}
In this work, we present our assessment of the state of the art \textit{feature-based}
interpretable and explainable ML for intrusion detection. In our study,
we highlight the concerning issues in ML-based intrusion detection research.
One such issue is the unjustified use of complex opaque DNNs 
while interpretable ML models, such as Ridge and DT classifiers, are powerful enough for
successful intrusion detection. Another key issue is the use of improper classification metrics. 
We empirically demonstrate that if proper metrics are not used, the standard metrics, such as
accuracy, precision, and F1-score, may overestimate the true performance and thus mislead the researchers.
Moreover, we show that stochastic uncertainties lead to inconsistent 
and unreliable \textit{feature-based explanations}. In addition, 
we find that strong feature correlations may exacerbate
this inconsistency problem. Furthermore, we introduce the novel method of \textit{cross-explanations} 
to better gauge if explanations are consistent and transferable. Applications of our method
corroborate that consistent and transferable explanations are absent for the datasets, the ML classifiers,
and the external explanation methods we study.
Finally, we find that hyper-parameters and optimization methods also cause unstable and inconsistent explanations.
Consequently, coupled with the lack of actionable feedback, feature-based 
explanations and the constructing methods do not offer convincing utility. 
Therefore, we advise that research 
should be focused on the approaches that are not based on the data features.

As future work, we plan to expand our study into more general ML settings
that are beyond intrusion detection and cybersecurity. 

\section*{Acknowledgments}
This work was supported by the U.S. DOE Office of Science, Office of Advanced Scientific Computing Research, under award 66150: "CENATE - Center for Advanced Architecture Evaluation" project. The Pacific Northwest National Laboratory is operated by Battelle for the U.S. Department of Energy under contract DE-AC05-76RL01830.







\section*{References} 

\bibliographystyle{unsrt}
\bibliography{references}

\begin{thebibliography}{10}

\bibitem{Goodfellow2016}
Ian Goodfellow, Yoshua Bengio, and Aaron Courville.
\newblock {\em Deep Learning}.
\newblock MIT Press, 2016.
\newblock {http://www.deeplearningbook.org}.

\bibitem{aisurvey1}
Ramanpreet Kaur, Dušan Gabrijelčič, and Tomaž Klobučar.
\newblock Artificial intelligence for cybersecurity: Literature review and future research directions.
\newblock {\em Information Fusion}, 97:101804, 2023.

\bibitem{macas2022survey}
Mayra Macas, Chunming Wu, and Walter Fuertes.
\newblock A survey on deep learning for cybersecurity: Progress, challenges, and opportunities.
\newblock {\em Computer Networks}, 212:109032, 2022.

\bibitem{molnar2020interpretable}
Christoph Molnar.
\newblock {\em Interpretable machine learning}.
\newblock Lulu. com, 2020.

\bibitem{surveyaicyber}
Gaith Rjoub, Jamal Bentahar, Omar Abdel~Wahab, Rabeb Mizouni, Alyssa Song, Robin Cohen, Hadi Otrok, and Azzam Mourad.
\newblock A survey on explainable artificial intelligence for cybersecurity.
\newblock {\em IEEE Transactions on Network and Service Management}, 20(4):5115--5140, 2023.

\bibitem{moustafa2023explainable}
Nour Moustafa, Nickolaos Koroniotis, Marwa Keshk, Albert~Y Zomaya, and Zahir Tari.
\newblock Explainable intrusion detection for cyber defences in the internet of things: Opportunities and solutions.
\newblock {\em IEEE Communications Surveys \& Tutorials}, 2023.

\bibitem{dodge}
Omer Subasi, Joseph Manzano, and Kevin Barker.
\newblock Denial of service attack detection via differential analysis of generalized entropy progressions.
\newblock In {\em 2023 IEEE International Conference on Cyber Security and Resilience (CSR)}, pages 219--226, 2023.

\bibitem{MATTHEWS1975442}
B.W. Matthews.
\newblock Comparison of the predicted and observed secondary structure of t4 phage lysozyme.
\newblock {\em Biochimica et Biophysica Acta (BBA) - Protein Structure}, 405(2):442--451, 1975.

\bibitem{Chicco2021-tv}
Davide Chicco, Niklas T{\"o}tsch, and Giuseppe Jurman.
\newblock The matthews correlation coefficient ({MCC}) is more reliable than balanced accuracy, bookmaker informedness, and markedness in two-class confusion matrix evaluation.
\newblock {\em BioData Min.}, 14(1):13, February 2021.

\bibitem{Chicco2023-la}
Davide Chicco and Giuseppe Jurman.
\newblock The matthews correlation coefficient ({MCC}) should replace the {ROC} {AUC} as the standard metric for assessing binary classification.
\newblock {\em BioData Min.}, 16(1):4, February 2023.

\bibitem{shapley}
Scott~M Lundberg and Su-In Lee.
\newblock A unified approach to interpreting model predictions.
\newblock In I.~Guyon, U.~V. Luxburg, S.~Bengio, H.~Wallach, R.~Fergus, S.~Vishwanathan, and R.~Garnett, editors, {\em Advances in Neural Information Processing Systems 30}, pages 4765--4774. Curran Associates, Inc., 2017.

\bibitem{breiman2001random}
Leo Breiman.
\newblock Random forests.
\newblock {\em Machine learning}, 45:5--32, 2001.

\bibitem{5gdataset}
Sehan Samarakoon, Yushan Siriwardhana, Pawani Porambage, Madhusanka Liyanage, Sang-Yoon Chang, Jinoh Kim, Jonghyun Kim, and Mika Ylianttila.
\newblock 5g-nidd: A comprehensive network intrusion detection dataset generated over 5g wireless network.
\newblock {\em IEEE Dataport}, 2022.

\bibitem{internetdataset}
Iman Sharafaldin, Arash~Habibi Lashkari, Saqib Hakak, and Ali~A. Ghorbani.
\newblock Developing realistic distributed denial of service (ddos) attack dataset and taxonomy.
\newblock In {\em 2019 International Carnahan Conference on Security Technology (ICCST)}, pages 1--8, 2019.

\bibitem{breiman2017classification}
Leo Breiman.
\newblock {\em Classification and regression trees}.
\newblock Routledge, 2017.

\bibitem{lime}
Marco~Tulio Ribeiro, Sameer Singh, and Carlos Guestrin.
\newblock "why should i trust you?": Explaining the predictions of any classifier.
\newblock In {\em Proceedings of the 22nd ACM SIGKDD International Conference on Knowledge Discovery and Data Mining}, KDD '16, page 1135–1144, New York, NY, USA, 2016. Association for Computing Machinery.

\bibitem{goldstein2015peeking}
Alex Goldstein, Adam Kapelner, Justin Bleich, and Emil Pitkin.
\newblock Peeking inside the black box: Visualizing statistical learning with plots of individual conditional expectation.
\newblock {\em journal of Computational and Graphical Statistics}, 24(1):44--65, 2015.

\bibitem{montavon2019layer}
Gr{\'e}goire Montavon, Alexander Binder, Sebastian Lapuschkin, Wojciech Samek, and Klaus-Robert M{\"u}ller.
\newblock Layer-wise relevance propagation: an overview.
\newblock {\em Explainable AI: interpreting, explaining and visualizing deep learning}, pages 193--209, 2019.

\bibitem{Prototype}
Jacob Bien and Robert Tibshirani.
\newblock {Prototype selection for interpretable classification}.
\newblock {\em The Annals of Applied Statistics}, 5(4):2403 -- 2424, 2011.

\bibitem{Bayesianproto}
Been Kim, Cynthia Rudin, and Julie Shah.
\newblock The bayesian case model: a generative approach for case-based reasoning and prototype classification.
\newblock In {\em Proceedings of the 27th International Conference on Neural Information Processing Systems - Volume 2}, NIPS'14, page 1952–1960, Cambridge, MA, USA, 2014. MIT Press.

\bibitem{prototypes2018}
Oscar Li, Hao Liu, Chaofan Chen, and Cynthia Rudin.
\newblock Deep learning for case-based reasoning through prototypes: a neural network that explains its predictions.
\newblock In {\em Proceedings of the Thirty-Second AAAI Conference on Artificial Intelligence and Thirtieth Innovative Applications of Artificial Intelligence Conference and Eighth AAAI Symposium on Educational Advances in Artificial Intelligence}. AAAI Press, 2018.

\bibitem{prototypenauta2021neural}
Meike Nauta, Ron Van~Bree, and Christin Seifert.
\newblock Neural prototype trees for interpretable fine-grained image recognition.
\newblock In {\em Proceedings of the IEEE/CVF conference on computer vision and pattern recognition}, pages 14933--14943, 2021.

\bibitem{karimi2020model}
Amir-Hossein Karimi, Gilles Barthe, Borja Balle, and Isabel Valera.
\newblock Model-agnostic counterfactual explanations for consequential decisions.
\newblock In {\em International conference on artificial intelligence and statistics}, pages 895--905. PMLR, 2020.

\bibitem{wachter2017counterfactual}
Sandra Wachter, Brent Mittelstadt, and Chris Russell.
\newblock Counterfactual explanations without opening the black box: Automated decisions and the gdpr.
\newblock {\em Harv. JL \& Tech.}, 31:841, 2017.

\bibitem{Anchors}
Marco~Tulio Ribeiro, Sameer Singh, and Carlos Guestrin.
\newblock Anchors: High-precision model-agnostic explanations.
\newblock {\em Proceedings of the AAAI Conference on Artificial Intelligence}, 32(1), Apr. 2018.

\bibitem{subasi2024analysis}
Omer Subasi, Sayan Ghosh, Joseph Manzano, Bruce Palmer, and Andr{\'e}s Marquez.
\newblock Analysis and benchmarking of feature reduction for classification under computational constraints.
\newblock {\em Machine Learning: Science and Technology}, 5(2):020501, 2024.

\bibitem{friedman2001greedy}
Jerome~H Friedman.
\newblock Greedy function approximation: a gradient boosting machine.
\newblock {\em Annals of statistics}, pages 1189--1232, 2001.

\bibitem{botnetdet}
Hatma Suryotrisongko, Yasuo Musashi, Akio Tsuneda, and Kenichi Sugitani.
\newblock Robust botnet dga detection: Blending xai and osint for cyber threat intelligence sharing.
\newblock {\em IEEE Access}, 10:34613--34624, 2022.

\bibitem{malwaredet}
Boryau Hsupeng, Kun-Wei Lee, Te-En Wei, and Shih-Hao Wang.
\newblock Explainable malware detection using predefined network flow.
\newblock In {\em 2022 24th International Conference on Advanced Communication Technology (ICACT)}, pages 27--33, 2022.

\bibitem{mane2021explaining}
Shraddha Mane and Dattaraj Rao.
\newblock Explaining network intrusion detection system using explainable ai framework.
\newblock {\em arXiv preprint arXiv:2103.07110}, 2021.

\bibitem{SHARMA2024121751}
Bhawana Sharma, Lokesh Sharma, Chhagan Lal, and Satyabrata Roy.
\newblock Explainable artificial intelligence for intrusion detection in iot networks: A deep learning based approach.
\newblock {\em Expert Systems with Applications}, 238:121751, 2024.

\bibitem{KESHK2023119000}
Marwa Keshk, Nickolaos Koroniotis, Nam Pham, Nour Moustafa, Benjamin Turnbull, and Albert~Y. Zomaya.
\newblock An explainable deep learning-enabled intrusion detection framework in iot networks.
\newblock {\em Information Sciences}, 639:119000, 2023.

\bibitem{data2017}
Iman Sharafaldin, Arash~Habibi Lashkari, Ali~A Ghorbani, et~al.
\newblock Toward generating a new intrusion detection dataset and intrusion traffic characterization.
\newblock {\em ICISSp}, 1:108--116, 2018.

\bibitem{nslkdd}
Mahbod Tavallaee, Ebrahim Bagheri, Wei Lu, and Ali~A. Ghorbani.
\newblock A detailed analysis of the kdd cup 99 data set.
\newblock In {\em 2009 IEEE Symposium on Computational Intelligence for Security and Defense Applications}, pages 1--6, 2009.

\bibitem{Sharafaldin2018TowardGA}
Iman Sharafaldin, Arash~Habibi Lashkari, and Ali~A. Ghorbani.
\newblock Toward generating a new intrusion detection dataset and intrusion traffic characterization.
\newblock In {\em International Conference on Information Systems Security and Privacy}, 2018.

\bibitem{TON_IoT}
Abdullah Alsaedi, Nour Moustafa, Zahir Tari, Abdun Mahmood, and Adnan Anwar.
\newblock Ton\_iot telemetry dataset: A new generation dataset of iot and iiot for data-driven intrusion detection systems.
\newblock {\em IEEE Access}, 8:165130--165150, 2020.

\bibitem{sarhan2022towards}
Mohanad Sarhan, Siamak Layeghy, and Marius Portmann.
\newblock Towards a standard feature set for network intrusion detection system datasets.
\newblock {\em Mobile networks and applications}, pages 1--14, 2022.

\bibitem{CIDDS-001}
Markus Ring, Sarah Wunderlich, Dominik Gr{\"u}dl, Dieter Landes, and Andreas Hotho.
\newblock Flow-based benchmark data sets for intrusion detection.
\newblock In {\em Proceedings of the 16th European Conference on Cyber Warfare and Security (ECCWS)}, pages 361--369. ACPI South Oxfordshire, UK, 2017.

\bibitem{CIDDS-002}
Markus Ring, Sarah Wunderlich, Dominik Gr{\"u}dl, Dieter Landes, and Andreas Hotho.
\newblock Creation of flow-based data sets for intrusion detection.
\newblock {\em Journal of Information Warfare}, 16(4):41--54, 2017.

\bibitem{UNSWNB15}
Nour Moustafa and Jill Slay.
\newblock Unsw-nb15: a comprehensive data set for network intrusion detection systems (unsw-nb15 network data set).
\newblock In {\em 2015 Military Communications and Information Systems Conference (MilCIS)}, pages 1--6, 2015.

\bibitem{samarakoon20225g}
Sehan Samarakoon, Yushan Siriwardhana, Pawani Porambage, Madhusanka Liyanage, Sang-Yoon Chang, Jinoh Kim, Jonghyun Kim, and Mika Ylianttila.
\newblock 5g-nidd: A comprehensive network intrusion detection dataset generated over 5g wireless network.
\newblock {\em arXiv preprint arXiv:2212.01298}, 2022.

\bibitem{Commercial_ids}
Camil Jichici, Bogdan Groza, Radu Ragobete, Pal-Stefan Murvay, and Tudor Andreica.
\newblock Effective intrusion detection and prevention for the commercial vehicle sae j1939 can bus.
\newblock {\em IEEE Transactions on Intelligent Transportation Systems}, 23(10):17425--17439, 2022.

\bibitem{BHARDWAJ2021100332}
Aanshi Bhardwaj, Veenu Mangat, Renu Vig, Subir Halder, and Mauro Conti.
\newblock Distributed denial of service attacks in cloud: State-of-the-art of scientific and commercial solutions.
\newblock {\em Computer Science Review}, 39:100332, 2021.

\bibitem{sklearn}
F.~Pedregosa, G.~Varoquaux, A.~Gramfort, V.~Michel, B.~Thirion, O.~Grisel, M.~Blondel, P.~Prettenhofer, R.~Weiss, V.~Dubourg, J.~Vanderplas, A.~Passos, D.~Cournapeau, M.~Brucher, M.~Perrot, and E.~Duchesnay.
\newblock Scikit-learn: Machine learning in {P}ython.
\newblock {\em Journal of Machine Learning Research}, 12:2825--2830, 2011.

\bibitem{tensorflow}
Martin Abadi, Ashish Agarwal, and Paul~Barham et. al.
\newblock {TensorFlow}: Large-scale machine learning on heterogeneous systems.
\newblock 2015.
\newblock https://www.tensorflow.org/.

\bibitem{keras}
Fran\c{c}ois Chollet et~al.
\newblock Keras.
\newblock 2015.
\newblock https://keras.io.

\bibitem{Mahbooba2021in}
Basim Mahbooba, Mohan Timilsina, Radhya Sahal, and Martin Serrano.
\newblock Explainable artificial intelligence ({XAI}) to enhance trust management in intrusion detection systems using decision tree model.
\newblock {\em Complexity}, 2021:6634811, January 2021.

\bibitem{dt-cidds-eval}
Maria~Camila Gaitan-Cardenas, Mahmoud Abdelsalam, and Kaushik Roy.
\newblock Explainable ai-based intrusion detection systems for cloud and iot.
\newblock In {\em 2023 32nd International Conference on Computer Communications and Networks (ICCCN)}, pages 1--7, 2023.

\bibitem{UNSW-NB15Analysis}
Shweta More, Moad Idrissi, Haitham Mahmoud, and A.~Taufiq Asyhari.
\newblock Enhanced intrusion detection systems performance with unsw-nb15 data analysis.
\newblock {\em Algorithms}, 17(2), 2024.

\bibitem{CICIoV2024ids}
Euclides Carlos~Pinto Neto, Hamideh Taslimasa, Sajjad Dadkhah, Shahrear Iqbal, Pulei Xiong, Taufiq Rahman, and Ali~A. Ghorbani.
\newblock Ciciov2024: Advancing realistic ids approaches against dos and spoofing attack in iov can bus.
\newblock {\em Internet of Things}, 26:101209, 2024.

\bibitem{CIC-IDS-2018study}
Surasit Songma, Theera Sathuphan, and Thanakorn Pamutha.
\newblock Optimizing intrusion detection systems in three phases on the cse-cic-ids-2018 dataset.
\newblock {\em Computers}, 12(12), 2023.

\bibitem{CIC-IDS2017study}
Arnaud ROSAY, Florent CARLIER, Elo\"{\i}se CHEVAL, and Pascal LEROUX.
\newblock From cic-ids2017 to lycos-ids2017: A corrected dataset for better performance.
\newblock In {\em IEEE/WIC/ACM International Conference on Web Intelligence and Intelligent Agent Technology}, WI-IAT '21, page 570–575, New York, NY, USA, 2022. Association for Computing Machinery.

\bibitem{NETO2024101209}
Euclides Carlos~Pinto Neto, Hamideh Taslimasa, Sajjad Dadkhah, Shahrear Iqbal, Pulei Xiong, Taufiq Rahman, and Ali~A. Ghorbani.
\newblock Ciciov2024: Advancing realistic ids approaches against dos and spoofing attack in iov can bus.
\newblock {\em Internet of Things}, 26:101209, 2024.

\bibitem{CICEV2023}
Yoonjib Kim, Saqib Hakak, and Ali Ghorbani.
\newblock Ddos attack dataset (cicev2023) against ev authentication in charging infrastructure.
\newblock In {\em 2023 20th Annual International Conference on Privacy, Security and Trust (PST)}, pages 1--9, 2023.

\bibitem{Imbalancedata}
Google Developers.
\newblock Imbalanced data, Accessed on 06-17-2024.
\newblock https://developers.google.com/machine-learning/data-prep/construct/sampling-splitting/imbalanced-data.

\bibitem{pml2Book}
Kevin~P. Murphy.
\newblock {\em Probabilistic Machine Learning: Advanced Topics}.
\newblock MIT Press, 2023.

\bibitem{sklearn-corr}
ScikitLearn Developers.
\newblock Permutation importance with multicollinear or correlated features.
\newblock Accessed on 05-17-2024.
\newblock https://scikit-learn.org/stable/auto$\_$examples/inspection/ plot$\_$permutation$\_$importance$\_$multicollinear.html\#sphx-glr-auto-examples-inspection-plot-permutation-importance-multicollinear-py.

\end{thebibliography}

\end{document}